\documentclass[journal, onecolumn, 12pt]{IEEEtran}

\usepackage{graphics}
\usepackage{rotating}
\usepackage{amsfonts}
\usepackage{amsmath}
\usepackage{subfigure}
\usepackage{verbatim}
\usepackage{enumerate}
\usepackage{setspace}
\usepackage{epsfig}

\graphicspath{{figures/}} \DeclareGraphicsExtensions{.eps,.ps}

\doublespace \centerfigcaptionstrue

\begin{document}

\title{Fairness Provision in the IEEE 802.11e Infrastructure Basic Service
Set \footnotemark{$^{\dag}$}}

\author{\singlespace \normalsize \authorblockN{Feyza Keceli, Inanc Inan, and Ender Ayanoglu}\\
\authorblockA{Center for Pervasive Communications and Computing \\
Department of Electrical Engineering and Computer Science\\
The Henry Samueli School of Engineering\\
University of California, Irvine, 92697-2625\\
%Phone: +1 949 824 9797  Fax : +1 949 824 2321\\
Email: \{fkeceli, iinan, ayanoglu\}@uci.edu}}

\maketitle

\footnotetext{$^{\dag}$ This work is supported by the Center for
Pervasive Communications and Computing, and by National Science
Foundation under Grant No. 0434928. Any opinions, findings, and
conclusions or recommendations expressed in this material are
those of authors and do not necessarily reflect the view of the
National Science Foundation.}

\begin{abstract}
Most of the deployed IEEE 802.11e Wireless Local Area Networks
(WLANs) use infrastructure Basic Service Set (BSS) in which an
Access Point (AP) serves as a gateway between wired and wireless
domains. We present the unfairness problem between the uplink and
the downlink flows of any Access Category (AC) in the 802.11e
Enhanced Distributed Channel Access (EDCA) when the default
settings of the EDCA parameters are used. We propose a simple
analytical model to calculate the EDCA parameter settings that
achieve weighted fair resource allocation for all uplink and
downlink flows. We also propose a simple model-assisted
measurement-based dynamic EDCA parameter adaptation algorithm.
Moreover, our dynamic solution addresses the differences in the
transport layer and the Medium Access Control (MAC) layer
interactions of  User Datagram Protocol (UDP) and Transmission
Control Protocol (TCP). We show that proposed Contention Window
(CW) and Transmit Opportunity (TXOP) limit adaptation at the AP
provides fair UDP and TCP access between uplink and downlink flows
of the same AC while preserving prioritization among ACs.
%We show that accurate
%uplink/downlink weighted fairness can ideally be achieved through
%a simple change in the EDCA backoff procedure at the AP. On the
%other hand, ideal condition assumptions of the theoretical
%analysis limit the practical use.
%To resolve the analytical inaccuracies that may occur as a result
%of non-ideal conditions in the WLAN,
\end{abstract}

\section{Introduction}

IEEE 802.11 Wireless Local Area Network (WLAN) is built around a
Basic Service Set (BSS) \cite{802.11}. While a number of stations
may gather to form an independent BSS with no connectivity to the
wired network, the common deployment is the infrastructure BSS
which includes an Access Point (AP). The AP provides the
connection to the wired network.

The IEEE 802.11 standard \cite{802.11} defines Distributed
Coordination Function (DCF) as a contention based Medium Access
Control (MAC) mechanism. The 802.11e standard \cite{802.11e}
updates the MAC layer of the former 802.11 standard for
Quality-of-Service (QoS) provision. In particular, the Enhanced
Distributed Channel Access (EDCA) function of 802.11e is an
enhancement of the DCF. The EDCA scheme (similarly to DCF) uses
Carrier Sense Multiple Access with Collision Avoidance (CSMA/CA)
and slotted Binary Exponential Backoff (BEB) mechanism as the
basic access method. The major enhancement to support QoS is that
EDCA differentiates packets using different priorities and maps
them to specific Access Categories (ACs) that use separate queues
at a station. Each AC$_{i}$ within a station ($0\leq i \leq 3$)
contends for the channel independently of the others. Levels of
services are provided through different assignments of the
AC-specific EDCA parameters; Contention Window (CW) sizes,
Arbitration Interframe Space (AIFS) values, and Transmit
Opportunity (TXOP) limits.

The DCF and the EDCA are defined such that each station in a BSS
uses the same contention parameter set. Therefore, fair access can
be achieved in the MAC layer for all the contending stations in
terms of the average number of granted transmissions, over a
sufficiently long interval. However, this does not translate into
achieving fair share of bandwidth between uplink and downlink
flows in the 802.11e infrastructure BSS. An AC of the AP which
serves all downlink flows has the same access priority with the
same AC of the stations that serve uplink flows. Therefore, an
approximately equal number of accesses that an uplink AC may get
is shared among all downlink flows in the same AC of the AP. This
leads to the uplink/downlink unfairness problem in the WLAN where
each individual downlink flow gets comparably lower bandwidth than
each individual uplink flow gets at high load. This phenomenon
will be described further in Section~\ref{sec:problemdefinition}.

%we deal with the weighted fair channel access
%support in the infrastructure IEEE 802.11e BSS.
We deal with weighted fair channel access between the uplink and
the downlink flows of the same AC in the IEEE 802.11e
infrastructure BSS. Using a simple analytical approach, we
calculate the EDCA parameter settings that achieve a given
utilization ratio between the uplink and the downlink
transmissions.
%that
%provides weighted fair channel access between uplink and downlink
%User Datagram Protocol (UDP) flows at high load. We show that
%accurate uplink/downlink weighted fairness can ideally be achieved
%through a simple change in the EDCA backoff procedure at the AP.
%The analytical calculation results in non-integer estimates
%of $CW_{min}$ values. Therefore, we also propose a simple change
%in the EDCA backoff procedure of the AP to approximate the
%performance of practical Binary Exponential Backoff (BEB)
%procedure with integer CW values to the analytical BEB model that
%uses non-integer CW values.
Comparing with simulation results, we noticed that sticking only
with analytical results that are based on ideal condition
assumptions may result in inaccuracies in a real WLAN scenario.
Therefore,
%we show that the theoretical decisions provide fair access among
%uplink/downlink real-time multimedia flows for ideal conditions.
we also propose a simple model-assisted measurement-based dynamic
EDCA parameter adaptation algorithm that provides weighted fair
resource allocation in an arbitrary scenario.

Most of the data traffic in the Internet is carried by
Transmission Control Protocol (TCP), while most of the real-time
applications use User Datagram Protocol (UDP). UDP employs one-way
unreliable communication. On the other hand, TCP defines reliable
bi-directional communication where the forward link data rate
depends on the rate of received Acknowledgment (ACK) packets in
the backward link. Another key contribution of this study is that
our solution considers the effects of this difference on the
design of the weighted fairness support algorithm.
%We show that a combination of the $CW_{min}$ adaptation with non-zero TXOP limit
%preserves weighted fair access among the uplink and the downlink
%TCP flows.

\section{Background}

In this section, we first present the uplink/downlink unfairness
problem in the IEEE 802.l1(e) WLAN at high traffic load. Next, we
provide a brief review of the literature on this subject.

%Due to space limitations, we do not include any overview on DCF,
%EDCA, UDP, and TCP, assuming that the reader has sufficient
%knowledge.

\subsection{Problem Definition} \label{sec:problemdefinition}

In the 802.11e WLAN, at high load, a bandwidth asymmetry exists
between contending upload and download flows which use the same
AC. This is due to the fact that the MAC layer contention
parameters are all equal for the AP and the stations. If $n$
stations and an AP are always contending for the access to the
wireless channel using the same AC, each host ends up having
approximately $1/(n+1)$ share of the total transmissions over a
long time interval. This results in $n/(n+1)$ of the transmissions
to be in the uplink, while only $1/(n+1)$ of the transmissions
belonging to the downlink flows. This is the WLAN uplink/downlink
unfairness problem stated previously. The uneven bandwidth share
results in downlink flows experiencing significantly lower
throughput and larger delay. The congestion at the AP may result
in considerable packet loss depending on the size of interface
buffers.

The results may even be more catastrophic in the case of TCP
flows.
% TCP flow and congestion control mechanisms run on
%bi-directional communication to ensure reliable transfer of the
%TCP data. The transmission data rate is adjusted according to the
%network capacity.
The TCP receiver returns TCP ACK packets to the TCP transmitter in
order to confirm the successful reception of data packets. In the
case of multiple uplink and downlink flows in the WLAN, returning
TCP ACKs of upstream TCP data are queued at the AP together with
the downstream TCP. When the bandwidth asymmetry in the forward
and reverse path builds up the queue in the AP, the dropped
packets impair the TCP flow and congestion control mechanisms
which assume equal transmission rate both in the forward and
reverse path \cite{Balakrishnan99}.

TCP's timeout mechanism initiates a retransmission of a data
packet if it has not been acknowledged during a \textit{timeout}
duration. However, any received TCP ACK can cumulatively
acknowledge all the data packets sent before the data packet for
which the ACK is intended to. When the packet loss is severe in
the AP buffer, downstream flows will experience frequent timeouts
resulting in significantly low throughput. On the other hand, due
to the cumulative property of TCP ACK mechanism, upstream flows
with high congestion windows will not experience such frequent
timeouts. In this case, it is a low probability that many
consecutive TCP ACK losses occur for the same flow. Conversely,
flows with low congestion window (fewer packets currently on
flight) may experience frequent timeouts and decrease their
congestion windows even more. Therefore, a number of upstream
flows may starve in terms of throughput while others enjoy a high
throughput. This results in unfairness between the TCP upstream
flows on top of the unfairness between the uplink and the
downlink.

\begin{figure}[t]
\centering \includegraphics[width = 1.0\linewidth]{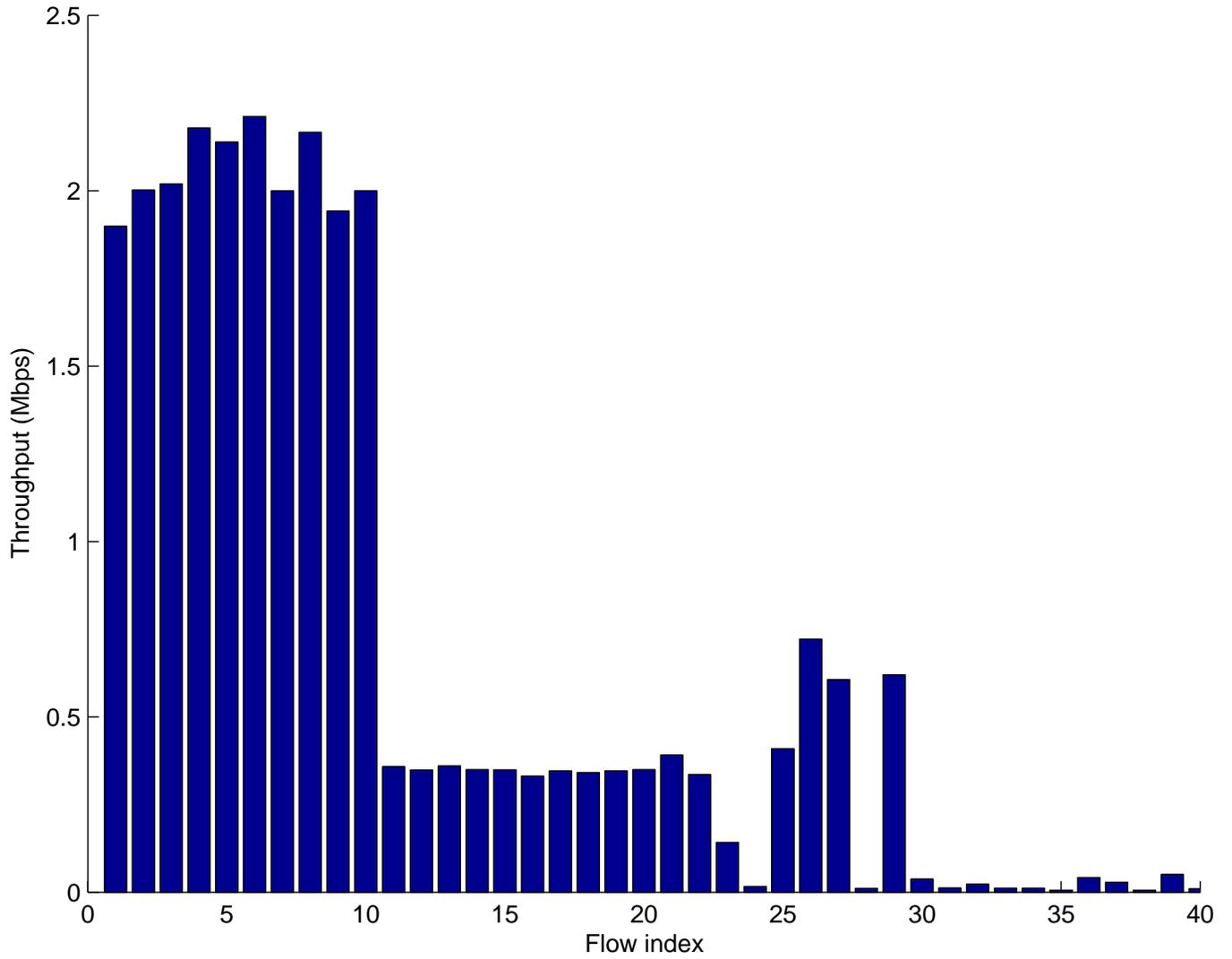}
\caption{Total throughput of 10 uplink UDP (indices 1-10), 10
downlink UDP (indices 11-20), 10 uplink TCP (indices 21-30) and 10
downlink TCP (indices 31-40 flows) when the AP and the stations
use equal EDCA parameters.} \label{fig:ind_thput}
\end{figure}

Fig.~\ref{fig:ind_thput} shows the average throughput of
individual flows for a scenario of 10 uplink UDP, 10 downlink UDP,
10 uplink TCP and 10 downlink TCP connections in an ns-2
simulation \cite{ns2},\cite{ourcode}. Each connection is initiated
by a separate station. All stations employ 54 Mbps data rate at
the physical layer. The packet size is 1500 bytes for all flows.
UDP flows are mapped to an AC with $CW_{min}=31$ and
$CW_{max}=511$. TCP flows use an AC with $CW_{min}=63$ and
$CW_{max}=1023$. For both ACs, AIFSN values are set to 2 and TXOP
limits are 0. Other simulation parameters are as stated in Section
\ref{sec:simulations}. The results illustrate the throughput
unfairness of the uplink and the downlink flows. The throughput
unfairness between uplink TCP connections is also significant.
Moreover, data packet losses at the AP buffer have almost shut
down all downlink TCP connections.

%Our main performance metric is the widely used fairness index
%\cite{Jain91}. The fairness index, $f$, is defined as follows: if
%there are $n$ concurrent connections in the network and the
%throughput achieved by connection $i$ is equal to $x_{i}$, $1 \leq
%i \leq n$, then
%\begin{equation}
%f =
%\frac{\left(\sum_{i=1}^{n}x_{i}\right)^{2}}{n\sum_{i=1}^{n}x_{i}^{2}}.
%\end{equation}
%
%The fairness index lies between 0 and 1, 1 being the most fair
%situation where every flow gets equal throughput. For the specific
%case illustrated in Fig.~\ref{fig:unfair}...

\subsection{Related Work}

There are two groups of studies in the literature related to this
work.

The first group works within the constraints of the default 802.11
contention parameters. In \cite{Pilosof03}, the effect of the AP
buffer size in the wireless channel bandwidth allocation for TCP
is studied. The proposed solution of \cite{Pilosof03} is to
manipulate advertised receiver windows of the TCP packets at the
AP. Uplink/downlink fairness problem is studied in \cite{Wu05}
using per-flow queueing. A simplified approach is proposed in
\cite{Ha06} where two separate queues for TCP data and ACKs are
used. In our previous work, we proposed using congestion control
and filtering techniques at the MAC layer to solve the TCP uplink
unfairness problem \cite{Keceli07_ICC}. Two queue management
strategies are proposed in \cite{Gong06} to improve TCP fairness.
A rate-limiter approach is used in \cite{Melazzi05} which requires
available instantaneous WLAN bandwidth estimation in both
directions.

The second group proposes changes at the MAC layer access
parameters to achieve improved fairness. Our work also falls into
this category. AIFS and CW differentiation is proposed for
improved fairness and channel utilization in \cite{Casetti04}. A
simulation-based analysis is carried out for a specific scenario
consisting of TCP and audio flows both in the uplink and the
downlink. An experimental study is carried out in \cite{Leith05}
to decide on CW and TXOP values of the AP and the stations for a
scenario with TCP uplink and downlink flows. Both solutions
propose that individual uplink and downlink streams use separate
ACs. No guidelines are provided on how to decide on the EDCA
parameters that achieve fair resource allocation for an arbitrary
scenario. Also, the interaction of TCP flow and congestion control
mechanisms with the MAC is not addressed. In \cite{SWKim05}, it is
proposed that the AP accesses the channel in Point Interframe
Space (PIFS) completion without any backoff when the interface
queue size goes over a threshold. The use of TXOP is evaluated in
\cite{Tinnirello05_2} for temporal fairness provisioning among
stations employing different data rates. Achieving weighted
fairness between uplink and downlink in DCF is studied through
mean backoff distribution adjustment in \cite{Jeong05}. A
mechanism that dynamically tunes CW and TXOP values in order to
prevent delay asymmetry of realtime UDP flows is proposed in
\cite{Freitag06}. An adaptive priority control mechanism is
employed in \cite{Shin06} to balance the uplink and downlink delay
of VoIP traffic.

%The problem of WLAN fairness has been analyzed at the MAC layer in
%\cite{Vaidya00}, \cite{Nandagopal00}.

\section{Weighted Fair Access between Uplink and Downlink Flows}

In this section, we first describe the simple analytical model we
propose in order to find the $AIFS$, $CW_{min}$, and $TXOP$
settings of the ACs that provide weighted fairness between uplink
and downlink flows.
%Since the calculated $CW_{min}$ values are not
%integers, we propose a practical extension of the BEB algorithm
%that uses integer $CW$ values to match the theoretical performance
%with non-integer $CW$ values.
%As most of the analytical studies in the literature, our analysis
%is accurate only for ideal conditions. The ideal condition
%assumption leads to analytical simplicity.
Next, we propose a parameter adaptation algorithm which
dynamically updates the analytically calculated CW and TXOP values
of the AP regarding simple network measurements. As we will
describe in Section~\ref{sec:tcpinter}, our dynamic solution also
addresses the effects of the slow-start phase of TCP.
% at the AP which converges to the appropriate $CW_{min}$ value in a couple of beacon intervals in
%the case of analytical inaccuracies at non-ideal conditions.
%We will also address the unfairness issues that may occur as a result of the
% flows which the studies in
%the literature have also missed. We show that assigning AP
%dynamically adjusted non-zero EDCA TXOPs overcomes this problem
%efficiently.

Every beacon interval, the AP announces the values of the
AC-specific EDCA parameters to the stations. The stations
overwrite their EDCA parameter settings with the new values if any
change is detected. Due to the specific design of the EDCA
Parameter Set element in the beacon packet, the stations can only
employ $CW$ values that are integer powers of 2, i.e., the AP
encodes the corresponding 4-bit fields of $CW_{min}$ and
$CW_{max}$ in an exponent form. A key point which the studies in
the literature have missed is that the CW settings of the ACs at
the AP are not restricted to the powers of 2. The ACs at the AP
may use any value and this value does not have to be equal to what
is announced via beacons.

\subsection{Analytical Model}\label{sec:analyticalmodel}

%Assuming slot homogeneity (constant collision and transmission
%probability at an arbitrary backoff slot) \cite{Bianchi00},
%\cite{Robinson04}, \cite{Hui05}, \cite{Inan07_ICC}, we model the
%BEB of the EDCA function for each traffic class. In the sequel, we
%use the terms, AC and Traffic Class (TC), interchangeably. As we
%will show,
Fair access between uplink and downlink flows using the same AC
can be provided by assigning different EDCA parameters for the AP
and the stations. This results in two Traffic Classes (TCs) using
the same AC. While uplink flows constitute the first TC, downlink
flows constitute the second TC. In the analysis, we will treat the
case with one AC (thus 2 TCs), since we address the weighted
fairness problem between the uplink and downlink flows that are
mapped to the same AC. Moreover, we only formulate the situation
when there is only one TC per station, therefore no internal
collisions can occur. Note that, this does not cause any loss of
generality, since the analysis can be extended for larger number
of ACs or TCs as in \cite{Inan07_cycle_trep}, and larger number of
ACs per station as in \cite{Kong04},\cite{Inan07_ICC}.

\begin{figure}
\center{\epsfig{file=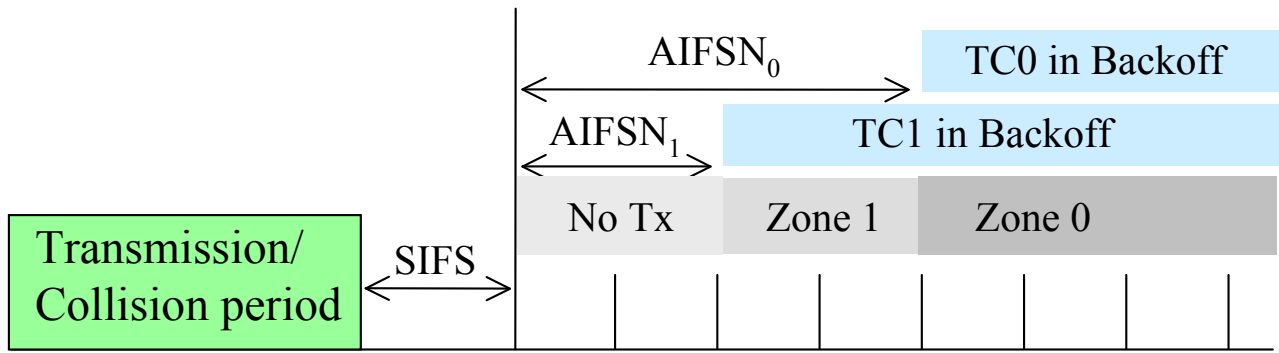,height=17 cm,angle=-90}} \caption[]
{\label{fig:contzones} EDCA backoff after busy medium. }
\end{figure}

Our analysis considers the fact that the difference in AIFS
creates the so-called contention zones as shown in
Fig.~\ref{fig:contzones}
\cite{Inan07_cycle_trep},\cite{Robinson04},\cite{Hui05},\cite{Inan07_ICC}.
First, we calculate the average collision probability of each TC
according to the long term occupancy of AIFS and backoff slots in
saturation. The average collision probability of a TC is a
function of transmission probabilities of all TCs. Next, we
formulate the average transmission probability for each TC, which
is a function of average collision probability of the same TC.
This results in a set of nonlinear equations which can be solved
numerically.

We define $p_{c_{i,x}}$ as the probability that TC$_{i}$
experiences a collision given that it has observed the medium idle
for $AIFS_{x}$ and transmits in the current slot (note
$AIFS_{x}\geq AIFS_{i}$ should hold). For notational simplicity,
let uplink flows belong to TC$_{0}$ and downlink flows belong to
TC$_{1}$. Let $d_{i} = AIFSN_{i} - AIFSN_{min}$ where
$AIFSN_{min}=\min(AIFSN_{0}, AIFSN_{1})$ and
$AIFS_{i}=SIFS+AIFSN_{i}\cdot T_{slot}$. Following the slot
homogeneity assumption of \cite{Bianchi00}, assume that each
TC$_{i}$ transmits with constant probability, $\tau_{i}$. Also,
let the total number of TC$_{i}$ in the BSS be $N_{i}$ (note that
$N_{1}=1$). Then,
\begin{equation}
\label{eq:pcix} \setlength{\nulldelimiterspace}{0pt} p_{c_{i,x}} =
1-\frac{\prod \limits_{i':d_{i'}\leq d_{x}}
(1-\tau_{i'})^{N_{i'}}}{(1-\tau_{i})}.
\end{equation}

\begin{figure}
\center{\epsfig{file=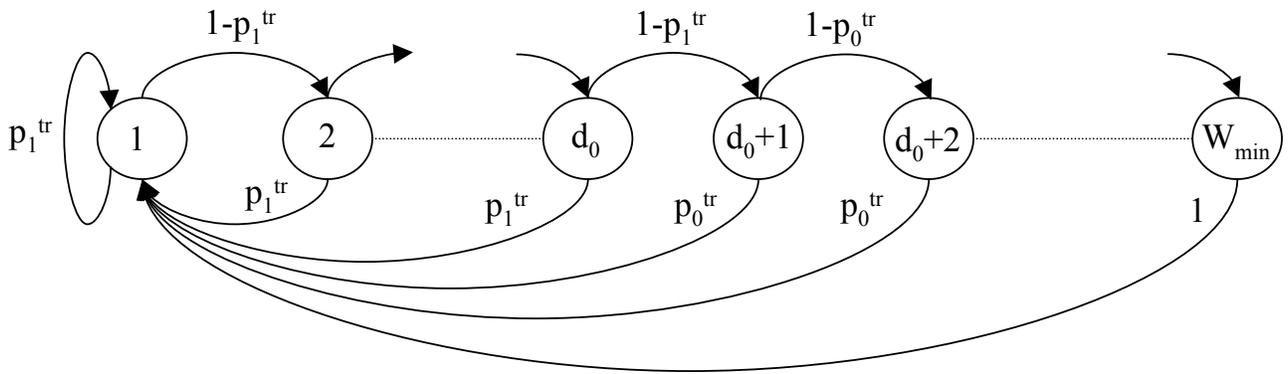,height=17 cm,angle=-90}}
\caption[] {\label{fig:AIFSMC} Transition through backoff slots in
different contention zones for the example given in
Fig.\ref{fig:contzones}.}
\end{figure}

We use the Markov chain shown in Fig.~\ref{fig:AIFSMC} to find the
long term occupancy of contention zones. Each state represents the
$n^{th}$ backoff slot after completion of the AIFS$_{min}$ idle
interval following a transmission period. The Markov analysis uses
the fact that a backoff slot is reached if no transmission occurs
in the previous slot. Moreover, the number of states is limited by
the maximum idle time between two successive transmissions which
is $W_{min}=\min(CW_{i,max})$ for a saturated scenario. The
probability that at least one transmission occurs in a backoff
slot in contention zone $x$ is
\begin{equation}
\label{eq:ptr} \setlength{\nulldelimiterspace}{0pt} p^{tr}_{x} =
1-\prod_{i':d_{i'}\leq d_{x}} (1-\tau_{i'})^{N_{i'}}.
\end{equation}

The long term occupancy of the backoff slots $b'_{n}$ in
Fig.~\ref{fig:AIFSMC} can be obtained from the steady-state
solution. Then, the average collision probability $p_{c_{i}}$ is
found by weighing zone specific collision probabilities
$p_{c_{i,x}}$ according to the long term occupancy of contention
zones (thus backoff slots)
\begin{equation}
\label{eq:pci}p_{c_{i}} = \frac{\sum_{n=d_{i}+1}^{W_{min}}
p_{c_{i,x}}b'_{n}}{\sum_{n=d_{i}+1}^{W_{min}} b'_{n}}
\end{equation}
\noindent where $x = \max \left( y~|~d_{y} = \underset{z}{\max}
(d_{z}~|~d_{z} \leq n)\right)$ which shows $x$ is assigned the
highest index value within a set of TCs that have AIFSN smaller
than equal to $n+AIFSN_{min}$.

%This ensures that at backoff slot $n$, TC$_{i}$ has observed the
%medium idle for AIFS$_{x}$. Therefore, the calculation
%in~(\ref{eq:pci}) fits into the definition of $p_{c_{i,x}}$.

Given $p_{c_{i}}$, we can calculate the expected number of backoff
slots $E_{i}[t_{bo}]$ that TC$_{i}$ waits before attempting a
transmission. Let $W_{i,k} = 2^{\min(k,m_{i})}(CW_{i,min}+1)-1$ be
the CW size of TC$_{i}$ at backoff stage $k$ where $CW_{i,max} =
2^{m_{i}}(CW_{i,min}+1)-1$, $0\leq m_{i} < r_{i}$. Note that, when
the retry limit $r_{i}$ is reached, any packet is discarded.
%Therefore, another $E_{i}[t_{bo}]$ passes between two
%transmissions with probability $p_{c_{i}}^{r_{i}}$.
\begin{equation}\label{eq:aveBO}
E_{i}[t_{bo}]=\sum_{n=0}^{\infty}(p_{c_{i}}^{r_{i}})^{n}\sum_{k=1}^{r}p_{c_{i}}^{k-1}(1-p_{c_{i}})\frac{W_{i,k}}{2}=\frac{1}{1-p_{c_{i}}^{r_{i}}}\sum_{k=1}^{r}p_{c_{i}}^{k-1}(1-p_{c_{i}})\frac{W_{i,k}}{2}.
\end{equation}

Then as also shown in \cite{Hui05}, the transmission probability
of TC$_{i}$ can be calculated as
\begin{equation}\label{eq:tauapp}
\tau_{i} = \frac{1}{E_{i}[t_{bo}]+1}.
\end{equation}

The nonlinear system of equations
(\ref{eq:pcix})-(\ref{eq:tauapp}) can be solved numerically to
calculate average collision and transmission probabilities of each
TC$_{i}$ for an arbitrary setting of EDCA parameters. We provide
the validation of the proposed analytical model in
\cite{Inan07_cycle_trep}.

\subsection{Weighted Fairness between Uplink and Downlink Flows}\label{sec:weightedfairness}

Let $\gamma_{i}$ be the probability that the transmitted packet
belongs to an arbitrary user from TC$_{i}$ given that the
transmission is successful. Also, let $p_{s_{i,n}}$ be the
probability that a successfully transmitted packet at backoff slot
$n$ belongs to AC$_{i}$. Then,
\begin{equation} \label{eq:gamma_i}
\gamma_{i} = \sum_{n=d_{i}+1}^{W_{min}}
b'_{n}\frac{p_{s_{i,n}}}{\sum \limits_{\forall j} p_{s_{j,n}}},
\end{equation}
\begin{equation}\label{eq:p_s_i_cycle}
p_{s_{i,n}} =
\left\{ \\
\begin{IEEEeqnarraybox}[\relax][c]{lc}
\frac{N_{i}\tau_{i}}{(1-\tau_{i})}\prod_{i':d_{i'}\leq
n-1}(1-\tau_{i'})^{N_{i'}}, &~{\rm if}~n \geq d_{i}+1 \\ 0, &~{\rm
if }~n < d_{i}+1.
\end{IEEEeqnarraybox}
\right.
\end{equation}

Let $U$ denote the utilization ratio between the downlink and the
uplink transmissions of an AC. Let $N_{TXOP,i}$ denote the maximum
number of packets that can fit in one TXOP of TC$_{i}$. Then, for
our running example with one AC,
\begin{equation}
\label{eq:U} U = \frac{\gamma_{1} \cdot N_{TXOP,1}}{\gamma_{0}
\cdot N_{TXOP,0}}.
\end{equation}

%If we let each TC$_{i}$ have an access weight of $w_{i}$,
%$\gamma_{i}/w_{i}$ should be equal for both traffic classes. Any
%combination of AIFS and CW values that satisfies this equality may
%be used.

 %In order to decrease
%the number of solutions that satisfies the same utilization ratio,

\subsubsection{Implementation of the Numerical Solution}

Without loss of generality, the EDCA parameters of the stations,
$AIFS_{0}$, $CW_{min,0}$, and $N_{TXOP,0}$, are fixed at
predetermined values . Then, the EDCA parameters of the TC at the
AP, $AIFS_{1}$, $CW_{min,1}$, and $N_{TXOP,1}$, that achieve a
required utilization ratio $U_{r}$ can be calculated numerically
as follows.

\begin{enumerate}
\item We assume AIFS differentiation is only used for the
prioritization between the ACs not the TCs (thus
$AIFS_{0}=AIFS_{1}$). \item When $AIFS_{0}=AIFS_{1}$, after some
algebra on (\ref{eq:gamma_i})-(\ref{eq:U}),
\begin{equation}
\label{eq:Usimp} U = \frac{\tau_{1}\cdot(1-\tau_{0})\cdot
N_{TXOP,1}}{\tau_{0}\cdot(1-\tau_{1})\cdot N_{TXOP,0}}.
\end{equation}
\noindent Therefore, $\tau_{1}$ can be written in terms of
$\tau_{0}$, $N_{TXOP,0}$, $N_{TXOP,1}$, and $U_{r}$. A numerical
solution for $\tau_{0}$ and $\tau_{1}$ for given $U_{r}$ and a
fixed value of $N_{TXOP,1}$ (initially, $N_{TXOP,1}=1$) is
obtained using (\ref{eq:pcix})-(\ref{eq:tauapp}). \item
$CW_{min,1}$ can be calculated as follows (the formula below is
obtained using (8) and (9) in \cite[Section IV-A]{Wu02}),
%\begin{equation}
%\label{eq:CWmin1} \setlength{\nulldelimiterspace}{0pt} CW_{min,i}
%=
%\left\{ \\
%\begin{IEEEeqnarraybox}[\relax][c]{ll} \frac{(2-\tau_{i})(1-p_{c_{i}}^{r_{i}})(1-2p_{c_{i}})(1-p_{c_{i}})}{\tau_{i}(1-p_{c_{i}})^{2}(1-(2p_{c_{i}})^{r_{i}})}-1, & ~ {\rm if}~m_{i} \geq r_{i} \\
%\frac{(2-\tau_{i})(1-p_{c_{i}}^{r_{i}})(1-2p_{c_{i}})(1-p_{c_{i}})}{\tau_{i}(1-p_{c_{i}})^{2}(1-(2p_{c_{i}})^{m_{i}+1})+\tau_{i}2^{m_{i}}p_{c_{i}}^{m_{i}+1}(1-2p_{c_{i}})(1-p_{c_{i}})(1-p_{c_{i}}^{r_{i}-m_{i}-1})}-1,
%& ~ {\rm if} ~ m_{i}<r_{i}.
%\end{IEEEeqnarraybox}
%\right.
%\end{equation}
\begin{equation}
\label{eq:CWmin1}CW_{min,i} =
\frac{2-\tau_{i}}{\tau_{i}}\cdot\frac{(1-p_{c_{i}}^{r_{i}})(1-2p_{c_{i}})(1-p_{c_{i}})}{(1-p_{c_{i}})^{2}(1-(2p_{c_{i}})^{m_{i}+1})+2^{m_{i}}p_{c_{i}}^{m_{i}+1}(1-2p_{c_{i}})(1-p_{c_{i}})(1-p_{c_{i}}^{r_{i}-m_{i}-1})}-1.
\end{equation}
\item A simple controller block checks whether the prioritization
among ACs are maintained or not for the new configuration. This
block ensures that $CW_{min}$ of a low priority AC (at the AP or a
station) is not smaller than $CW_{min}$ of a higher priority AC.
Therefore, if analytically calculated $CW_{min,1}$ value does not
satisfy the controller block requirements, $N_{TXOP,1}$ is doubled
and the algorithm returns to step 3. The larger $N_{TXOP,1}$ is,
the larger $CW_{min,1}$ will be. \item If the calculated
$CW_{min,1}$ is not an integer, it is rounded to the closest
integer value.
\end{enumerate}

A few remarks on the implementation are as follows.
\begin{itemize}
\item A numerical solution also exists when $AIFS_{0}$ and
$AIFS_{1}$ are not equal, but the implementation differs since
(\ref{eq:Usimp}) does not hold. In such a case, $AIFS_{1}$ is also
assigned an initial value as $N_{TXOP,1}$ and the nonlinear system
of equations (\ref{eq:pcix})-(\ref{eq:U}) is solved numerically.
According to the controller block requirements on $CW_{min,1}$,
the procedure may be repeated for updated values of $AIFS_{1}$ and
$N_{TXOP,1}$. \item As previously mentioned, our formulation is
valid for the situation when there is only one TC per station
(including the AP). As an approximation, we assume that
(\ref{eq:pcix})-(\ref{eq:U}) still holds when there are multiple
TCs at the AP. Indeed as only a few collisions are avoided when
the internal collision procedure is run at the AP \cite{Banchs06},
the solution of (\ref{eq:pcix})-(\ref{eq:U}) will be very close to
an extension that exactly formulates the virtual collisions at the
AP. In this case, if the AIFS values of TCs within an AC remains
equal, it can be shown that (\ref{eq:Usimp}) still holds for the
TCs of the same AC. Therefore, we use the implementation procedure
previously stated for scenarios when larger number of ACs exist as
long as there is one AC (or TC) per station and multiple TCs at
the AP.
\end{itemize}

\subsubsection{Proposed BEB Algorithm for non-integer CW
values}\label{sec:BEBalgorithm}

As specified in \cite{802.11e}, the initial value of $CW$ is set
to the AC-specific $CW_{min}$. At each unsuccessful transmission,
the value of $CW$ is doubled until the maximum AC-specific
$CW_{max}$ limit is reached. The value of $CW$ is reset to the
AC-specific $CW_{min}$ if the transmission is successful, or the
retry limit is reached thus the packet is dropped.

The proposed analytical calculation for weighted fairness may
decide a non-integer value of $CW_{min,1}$ thus $W_{1,k}$,
$k<r_{1}$. The simplest approach is rounding to the closest
integer and employing the rounded value in the BEB.

Instead, we also propose the AP to choose integer $W'_{1,k}$
values from a probability distribution that satisfies $E[W'_{1,k}]
= W_{1,k}$. For example, it is straightforward to show a simple
discrete probability distribution such as ${\rm Pr}(W'_{1,k} =
\lfloor W_{1,k} \rfloor) = \lceil W_{1,k} \rceil-W_{1,k}$ and
${\rm Pr}(W'_{1,k} = \lceil W_{1,k} \rceil) = W_{1,k}-\lfloor
W_{1,k} \rfloor$ holds. According to the proposed algorithm, the
EDCA function at the AP decides on the interval $(0,W'_{1,k})$ to
select the backoff value regarding the given simple discrete
probability distribution.

\begin{figure}[t]
\centering \includegraphics[width = 1.0\linewidth]{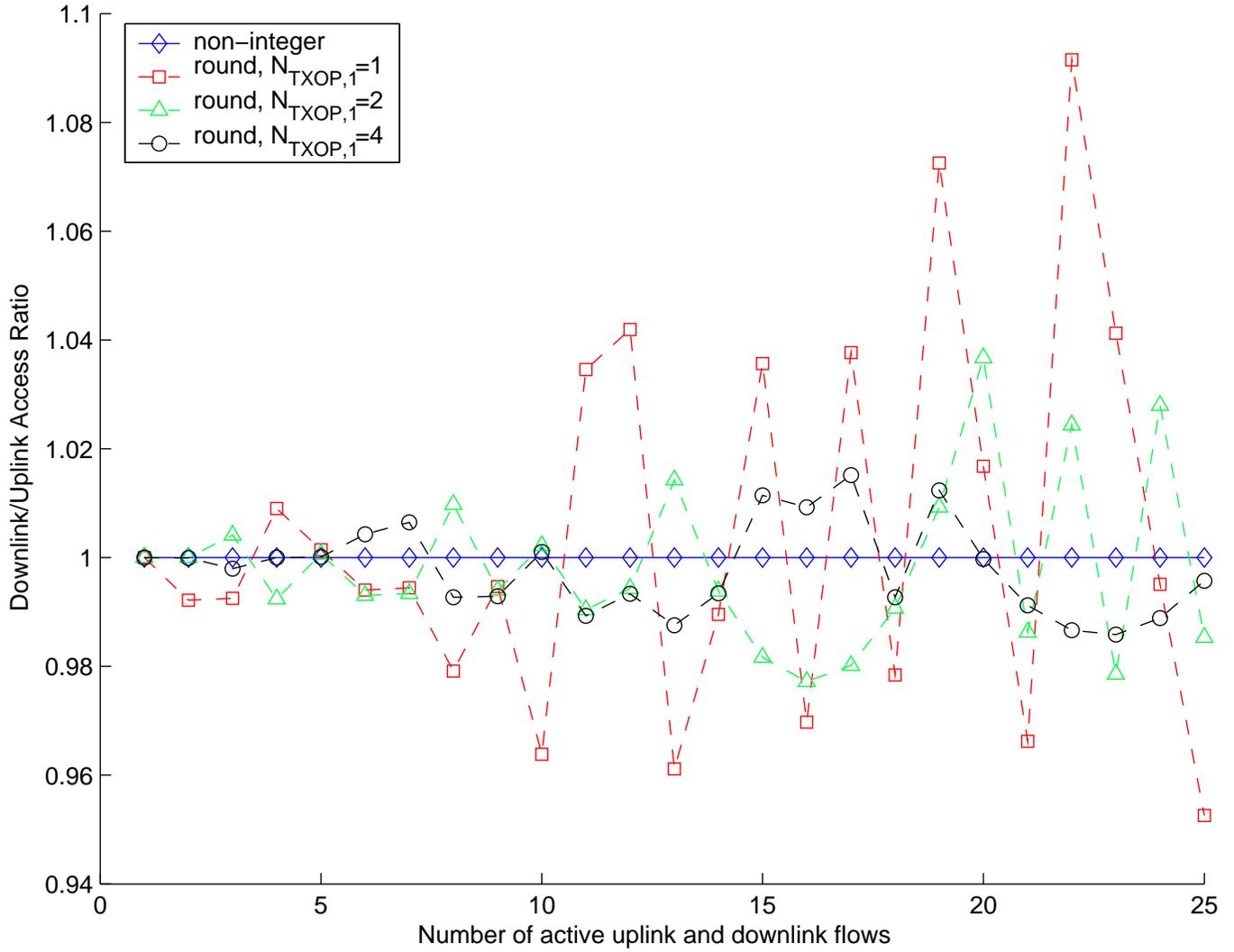}
\caption{The downlink/uplink access ratio for increasing number of
uplink and downlink flows.} \label{fig:BEB}
\end{figure}

Fig. \ref{fig:BEB} shows the downlink/uplink access ratio for
increasing number of uplink and downlink flows. We assume equal
$AIFSN=2$ for all the stations and the AP, and analytically
calculate $CW_{min,1}$ that achieves downlink/uplink access ratio
of $U_{r}=1$ when $CW_{min,0}=127$, $N_{TXOP,0}=1$, and
$N_{TXOP,1}$ is varied from 1 to 4. The performance of rounding
the analytically calculated CW values is compared with the
performance of the proposed BEB algorithm that uses the stated
discrete probability distribution function. As the results imply,
the proposed BEB algorithm maintains perfect weighted fairness
while rounding the analytically calculated value may result in
slight inaccuracies in terms of utilization ratio. As the number
of uplink stations increase, $CW_{min,1}$ that achieves $U_{r}=1$
decreases. As Fig. \ref{fig:BEB} shows the effect of rounding is
much more noticeable when $CW_{min,1}$ is small. The effect of
rounding becomes negligible as $N_{TXOP,1}$ (thus $CW_{min,1}$) is
increased.

\subsection{Dynamic Parameter Adaptation}\label{sec:dynalgorithm}

The IEEE 802.11 infrastructure BSS exhibits some non-ideal
conditions which most of the analytical models ignore to maintain
simplicity. For example,
\begin{itemize}
\item
%The proposed model uses the instantaneous number of active
%stations in the calculation. However,
Accurate information on the instantaneous number of active flows
may not always be available to the AP \cite{Bianchi03}.
%The
%802.11e standard requires each multimedia flow to run a QoS
%reservation procedure with the AP for QoS provision and admission
%control.
%Therefore, the
%information on the number of active multimedia flows can be
%readily available at the AP.
%On the other hand, such a restriction
%does not apply to best-effort data flows.
%The approach in
%\cite{Bianchi03} can be used to estimate the number of active data
%stations.
\item
%We have observed that even with the theoretically fair EDCA
%settings, the AP receives higher throughput than the stations in
%some simulation scenarios. Note that
If a station and the AP collide, the station's transmission
results in failure since the destination (the AP) is not in listen
mode. However, there is some probability that the transmission of
the AP results in success as a consequence of the capture effect
depending on the spatial distribution and the power levels of the
stations \cite{JKLee99}.
\end{itemize}

Such non-ideal conditions make finding the optimum EDCA setting
analytically hard for any scenario. This also limits the use of
proposed BEB algorithm for non-integer CW values. We propose a
simple model-assisted measurement-based dynamic algorithm to adapt
the analytically calculated $CW_{min}$ values for such scenarios.

The AP carries out the dynamic adaptation for each AC every
$\beta$ beacon intervals which is called an \textit{adaptation
interval} in the sequel. If it is detected as a new flow starting
transmission or as an old flow becoming inactive at the last
adaptation interval, the algorithm decides on new \textit{good}
EDCA parameters using the proposed analytical model which results
in weighted fair resource allocation for the estimated number of
uplink and downlink flows in ideal conditions. Otherwise, fine
tuning on the $CW$ and the $TXOP$ values of the AC at the AP is
carried out to make measured $U$ as close as to $U_{r}$.

We use a simple algorithm to estimate the number of active flows.
More advanced approaches \cite{Bianchi03} can also be used. The AP
counts the number of unique source and destination MAC addresses
observed from incoming frames to estimate the number of uplink and
downlink flows respectively. Let $n_{u}$ and $n_{d}$ denote the
number of uplink and downlink flows labeled as active. If the AP
receives a packet with the corresponding MAC address not on its
list, it adds the new MAC address to the list and increments
$n_{u}$ or $n_{d}$. If the AP does not receive any packet with the
corresponding MAC address during the last adaptation interval, it
deletes the MAC address from the list and decrements $n_{u}$ or
$n_{d}$. Then, we define the required utilization ratio as
\begin{equation}
\label{eq:Ur} U_{r}=\frac{n_{d}}{n_{u}}.
\end{equation}

If $U_{r}$ has been changed during the last adaptation interval,
EDCA parameters are analytically calculated for $U=U_{r}$ and the
fine tuning phase is skipped. Otherwise, solely fine tuning on
$CW_{min}$ is performed as follows. Every adaptation interval, the
AP measures the number of successful uplink and downlink
transmissions, $n_{t_{u}}$ and $n_{t_{d}}$ respectively where
$n_{t_{d}}/n_{t_{u}}$ is the measured $U$ of the last adaptation
interval. If $\frac{n_{t_{d}}}{n_{t_{u}}} < (1-\alpha) \cdot
U_{r}$, then $CW_{min,1}$ is decremented (where $0\leq \alpha \leq
1$). Similarly, if $\frac{n_{t_{d}}}{n_{t_{u}}}
> (1+\alpha) \cdot U_{r}$, then $CW_{min,1}$ is incremented.
Otherwise, no action is taken. Note that using steps equal in
value to 1 in the $CW_{min}$ adaptation is sufficient since the
analytical calculation will provide a good initial guess.
%Such fine tuning removes additional analytical inaccuracies if they
%exist.

%A controller block should also check whether the prioritization
%among ACs are maintained or not. For example, $CW_{min}$ of a low
%priority AC (at the AP or a station) should not be smaller than
%$CW_{min}$ of a higher priority AC. Since the analytical
%calculation fixes the station parameters and calculates the AP
%settings accordingly, an appropriate initial selection for station
%parameters can result in AP settings that maintain prioritization
%between ACs.

\subsection{TCP-MAC Interactions}\label{sec:tcpinter}

TCP defines a reliable bi-directional communication where the
forward link data rate depends on the rate of the received ACK
packets in the backward link. This behavior of TCP constitutes the
main difference between TCP and UDP access in the WLAN. The key
observation is that, if we assume there are no packet losses in
TCP connections (infinitely large interface buffers at the AP and
the stations), the TCP access is fair irrespective of the EDCA
parameter selection (which is not the case for UDP). This is due
to the fact that the slow link limits the throughput for all TCP
flows. However, when the buffer size at the AP (bottleneck) is
limited, significant unfairness and low channel utilization is
experienced as previously shown in Fig.~\ref{fig:ind_thput}.
Therefore, for fair resource allocation and high channel
utilization, packet losses at the AP buffer should be minimized.
%The studies in the literature try to achieve
%this objective using different strategies.
We configure our adaptation algorithm considering the TCP dynamics
to achieve this objective.

None of the work in the literature on IEEE 802.11 MAC
upload/download fairness considered the asymmetry in the forward
and backward link packet rate during the slow-start phase of the
TCP connections. During the slow-start phase, the packet rate in
the forward link is twice the packet rate in the backward link.
When the congestion avoidance phase is entered, the forward and
the backward link packet rates become equal. When this asymmetry
during slow-start is neglected, the download traffic is penalized
with longer queueing delays.
%Neglecting the effects of slow-start results in long packet delays in the AP
%buffer.
Depending on the buffer availability, significant packet
loss may even occur during the slow-start. These may considerably
affect the short-term fairness and the channel utilization.

Our solution is simple yet effective. Considering each TCP data
and ACK streams of each connection as individual active flows, the
parameter adaptation algorithm of Section \ref{sec:dynalgorithm}
is used. Since TCP is fair irrespective of the EDCA parameter
selection as long as there are no packet losses, fine tuning on
$CW_{min}$ is always skipped. Therefore, the AP does not have to
measure $n_{t_{u}}$ and $n_{t_{d}}$. On the other hand, fine
tuning is carried out on TXOP assignments to overcome increased
rate of downlink TCP data flows during slow-start. Since the
forward to backward link packet rate ratio is 2 during the
slow-start, the analytically calculated TXOP duration is
multiplied by 2. Our approach is adapting the duration of the TXOP
depending on the number of packets buffered at the interface
queue. If the number of packets goes over a threshold value $th$,
doubled TXOPs are enabled until the number goes below the
threshold again.
%Since any unused portion of EDCA TXOP is
%returned, the AP may always stick with fixed duration EDCA TXOPs.
%On the other hand, such an approach may result in marginally
%higher downlink weight.

Assigning best-effort data flows a non-zero TXOP or a small
$CW_{min}$ may not be a favorable approach when multimedia flows
coexist in the WLAN. The controller block located at the AP should
check whether the QoS for admitted realtime flows is preserved or
not in the WLAN with the $CW_{min}$ and $TXOP$ values calculated
for uplink/downlink fairness.

%On the other hand, changing these values only for one station
%(only the AP according to the proposed approach) does not
%considerably affect the QoS of existing flows as we will show in
%Section \ref{sec:simulations}.

\section{Numerical and Simulation Results}\label{sec:simulations}

We carried out simulations in ns-2 \cite{ns2} in order to evaluate
the performance of the proposed weighted fairness adaptation
algorithm. For the simulations, we employ the IEEE 802.11e EDCA
MAC simulation module for ns-2.28 \cite{ourcode}.

We consider a network topology where each wireless station
initiates a connection with a wired station where the WLAN traffic
is relayed to the wired network through the AP. The stations are
uniformly distributed on a circle and the AP is located at the
center. The power thresholds are set so that every station can
hear the other's transmission. The data connections use either UDP
or TCP NewReno. The UDP traffic uses a Constant Bit Rate (CBR)
application. The TCP traffic uses a File Transfer Protocol (FTP)
agent which models bulk data transfer. The default TCP NewReno
parameters in ns-2 are used. The UDP traffic is mapped to a higher
priority AC than the TCP traffic. All the stations are assumed to
have 802.11g PHY using 54 Mbps and 6 Mbps as the data and basic
rate respectively \cite{802.11g}. The packet size is 1500 bytes
for all flows. The buffer size at the stations and the AP is set
to 200 packets. We found $\beta=5$, $\alpha=0.5$, and $th=50$
packets to be appropriate through extensive simulations.

\begin{figure}[t]
\centering \includegraphics[width = 1.0\linewidth]{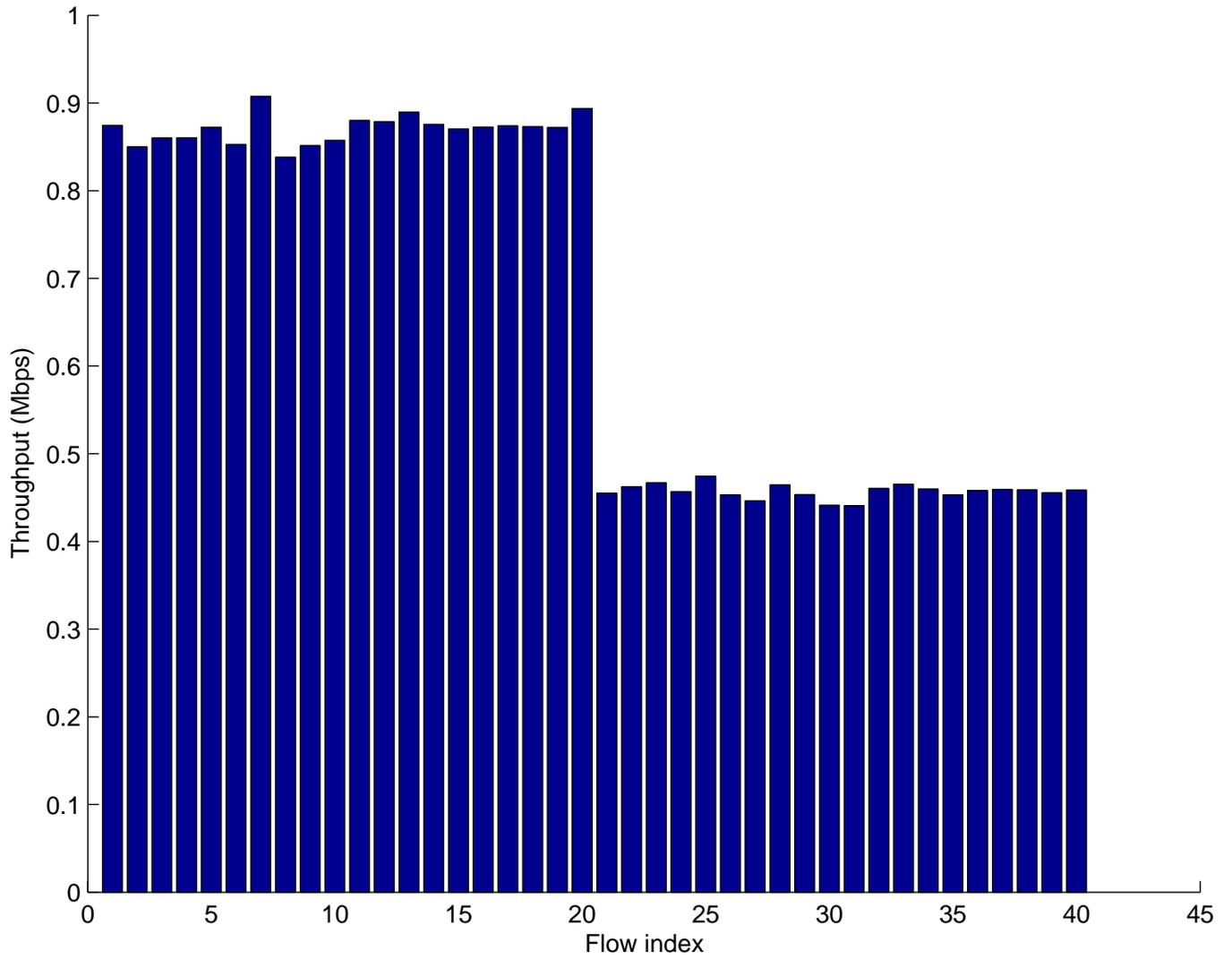}
\caption{Total throughput of 10 uplink UDP (indices 1-10), 10
downlink UDP (indices 11-20), 10 uplink TCP (indices 21-30) and 10
downlink TCP (indices 31-40 flows) when the AP uses the proposed
adaptation algorithm to achieve $U_{r}=1$.}
\label{fig:ind_thput_fair}
\end{figure}

Fig.~\ref{fig:ind_thput_fair} shows the average throughput of
individual flows for a scenario of 10 uplink UDP, 10 downlink UDP,
10 uplink TCP and 10 downlink TCP connections (same scenario as in
Fig.~\ref{fig:ind_thput}). At the stations, UDP flows are mapped
to an AC with $CW_{min}=31$ and $CW_{max}=511$. TCP flows use an
AC with $CW_{min}=63$ and $CW_{max}=1023$. For both ACs, AIFSN
values are set to 2 and TXOP limits are 0. Unless otherwise
stated, all data connections of the stations in other experiments
use these ACs (thus these EDCA parameters). At the AP, we run the
proposed algorithm designed for weighted fairness support in the
downlink and uplink. Since the number of downlink and uplink flows
are equal for both ACs, we define the downlink/uplink utilization
requirement as $U_{r}=1$.
%We set $N_{TXOP}$ to 2 and 4 for UDP flows respectively.
The analytical model decides on the $CW$ and the $TXOP$ that
achieves $U_{r}=1$. Fine tuning on $CW$ is carried out for the
fairness of UDP flows. The $TXOP$ is adaptively doubled according
to the proposed algorithm for TCP flows. The results illustrate
that $U=1$ is perfectly achieved in terms of throughput for both
UDP and TCP flows.

We have tested the proposed algorithm for a range of network
conditions.

\begin{figure}[t]
\centering \includegraphics[width = 1.0\linewidth]{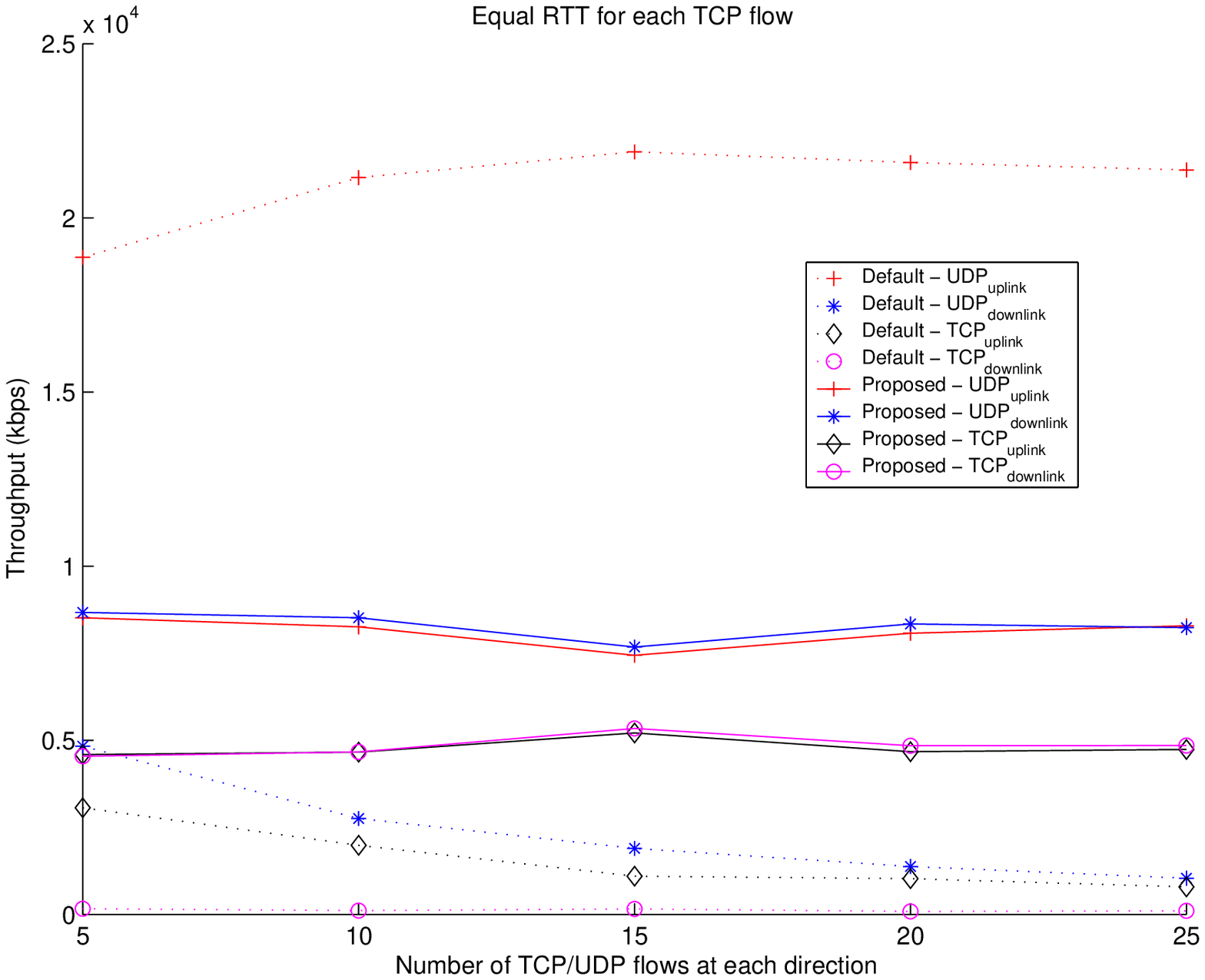}
\caption{The total throughput of TCP and UDP flows in each
direction (experiment 1).} \label{fig:udp_sc1_fig3}
\end{figure}

\begin{figure}[t]
\centering \includegraphics[width = 1.0\linewidth]{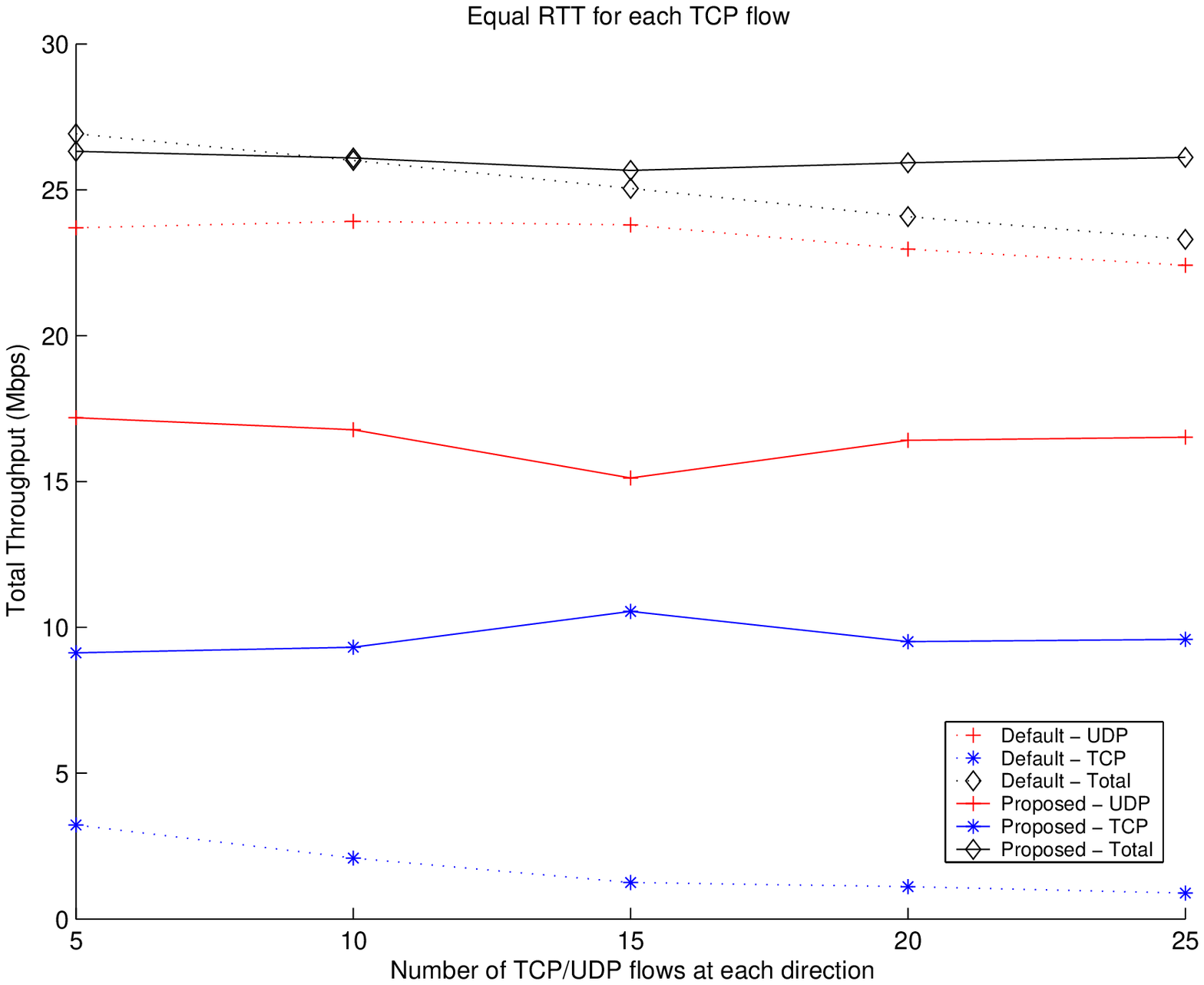}
\caption{The total throughput of TCP and UDP flows as well as the
total system throughput (experiment 1).} \label{fig:udp_sc1_fig5}
\end{figure}

\begin{figure}[t]
\centering \includegraphics[width = 1.0\linewidth]{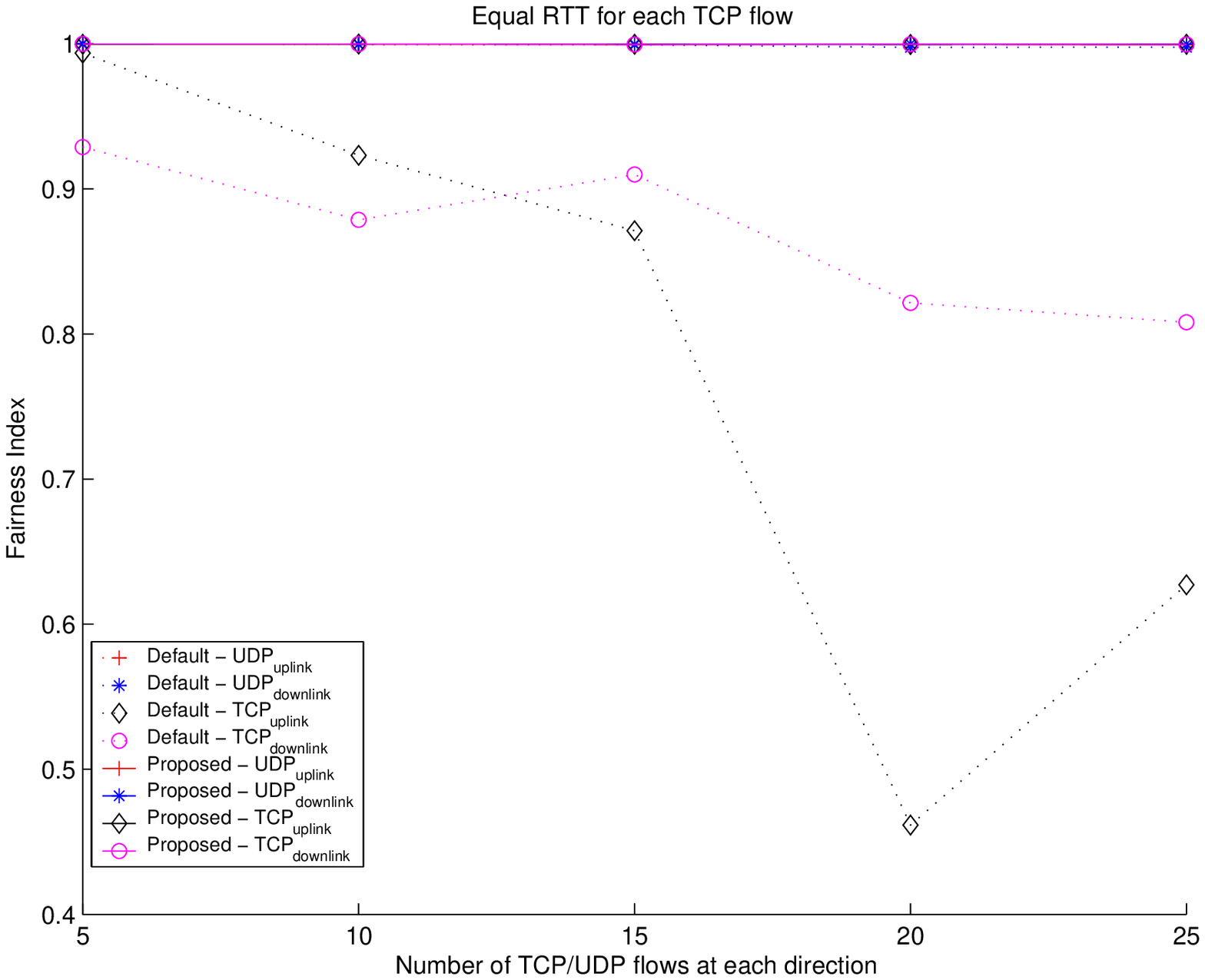}
\caption{Fairness index of individual TCP or UDP flows in the same
direction (experiment 1).} \label{fig:udp_sc1_fig1}
\end{figure}

\paragraph{Experiment 1}
In the first set of experiments, we generate an equal number of
TCP and UDP flows both in the uplink and downlink. Each flow
starts at the same time and the simulation duration is 100
seconds. The wired link delay (denoted as Round Trip Time (RTT) in
the titles of the figures) is equal for all flows (30 ms).
Fig.~\ref{fig:udp_sc1_fig3} shows the total throughput of TCP and
UDP flows in each direction for the proposed algorithm. The
results for the default 802.11e EDCA are also included for
comparison. As the results depict, $U=1$ is perfectly achieved in
terms of average throughput for the proposed algorithm, while the
default scheduler cannot maintain fair access.
Fig.~\ref{fig:udp_sc1_fig5} shows the total throughput of TCP and
UDP flows as well as the total system throughput for the proposed
algorithm and the default case. The proposed algorithm can
maintain more efficient channel utilization than the default EDCA
while providing fair access. In Fig.~\ref{fig:udp_sc1_fig1}, we
present the performance in terms of fairness between individual
TCP or UDP flows in the same direction for the proposed algorithm
and the default EDCA. The performance metric we use is the widely
used fairness index \cite{Jain91}. The fairness index, $f$, is
defined as follows: if there are $n$ concurrent connections and
the throughput achieved by connection $i$ is equal to $x_{i}$, $1
\leq i \leq n$, then
\begin{equation}
f =
\frac{\left(\sum_{i=1}^{n}x_{i}\right)^{2}}{n\sum_{i=1}^{n}x_{i}^{2}}.
\end{equation}
\noindent As the results imply, the proposed algorithm also
provides fair access between UDP and TCP flows of the same
direction. However, the default EDCA results in unfair resource
allocation even between the TCP flows of the same direction. As we
have described in Section \ref{sec:problemdefinition}, the
unfairness is more significant between TCP uplink flows. Although
no unfair behavior is expected between UDP flows in the same
direction, we have included these results in
Fig.~\ref{fig:udp_sc1_fig1} for the sake of completeness.

\begin{figure}[t]
\centering \includegraphics[width = 1.0\linewidth]{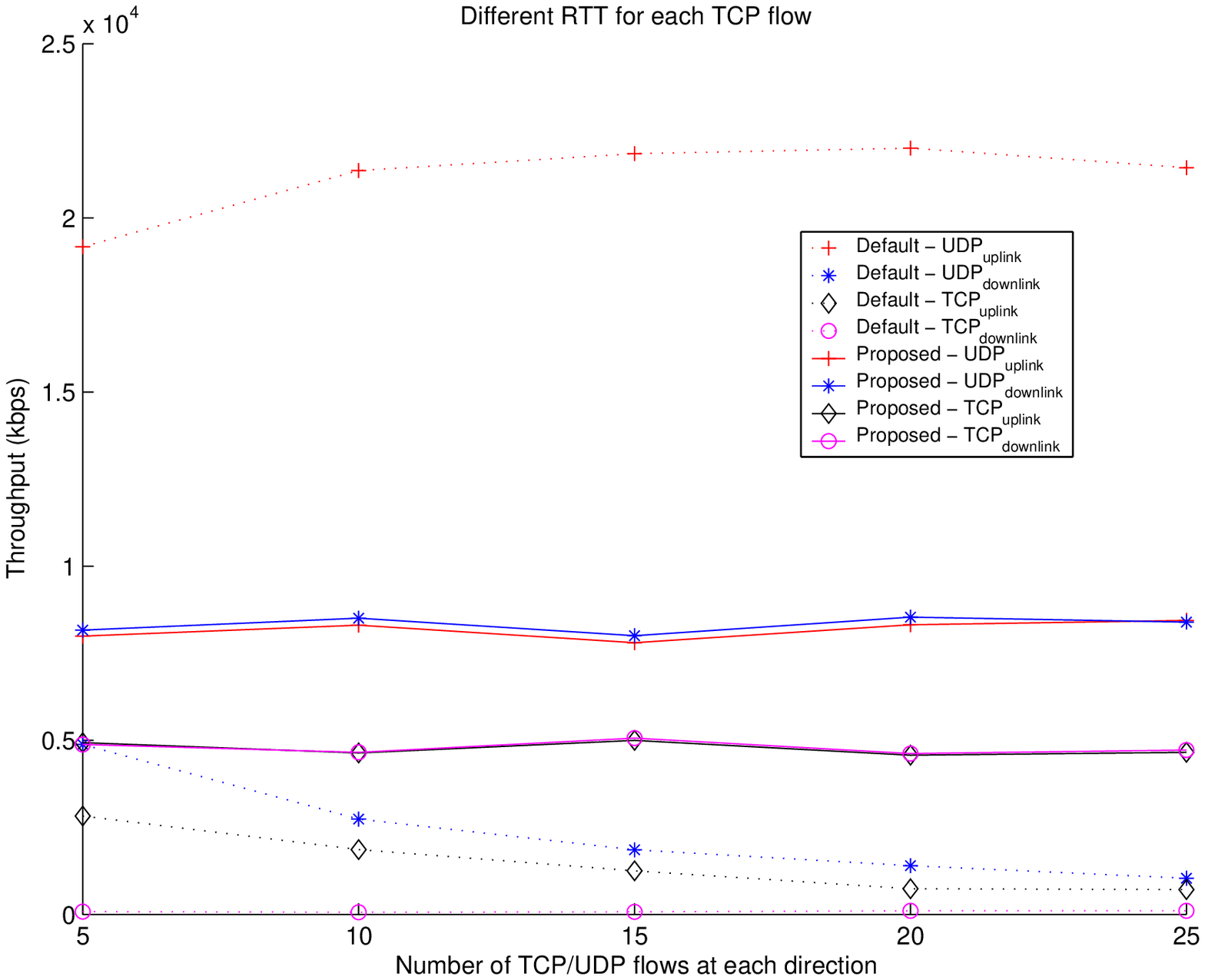}
\caption{The total throughput of TCP and UDP flows in each
direction (experiment 2).} \label{fig:udp_sc1_fig4}
\end{figure}

\begin{figure}[t]
\centering \includegraphics[width = 1.0\linewidth]{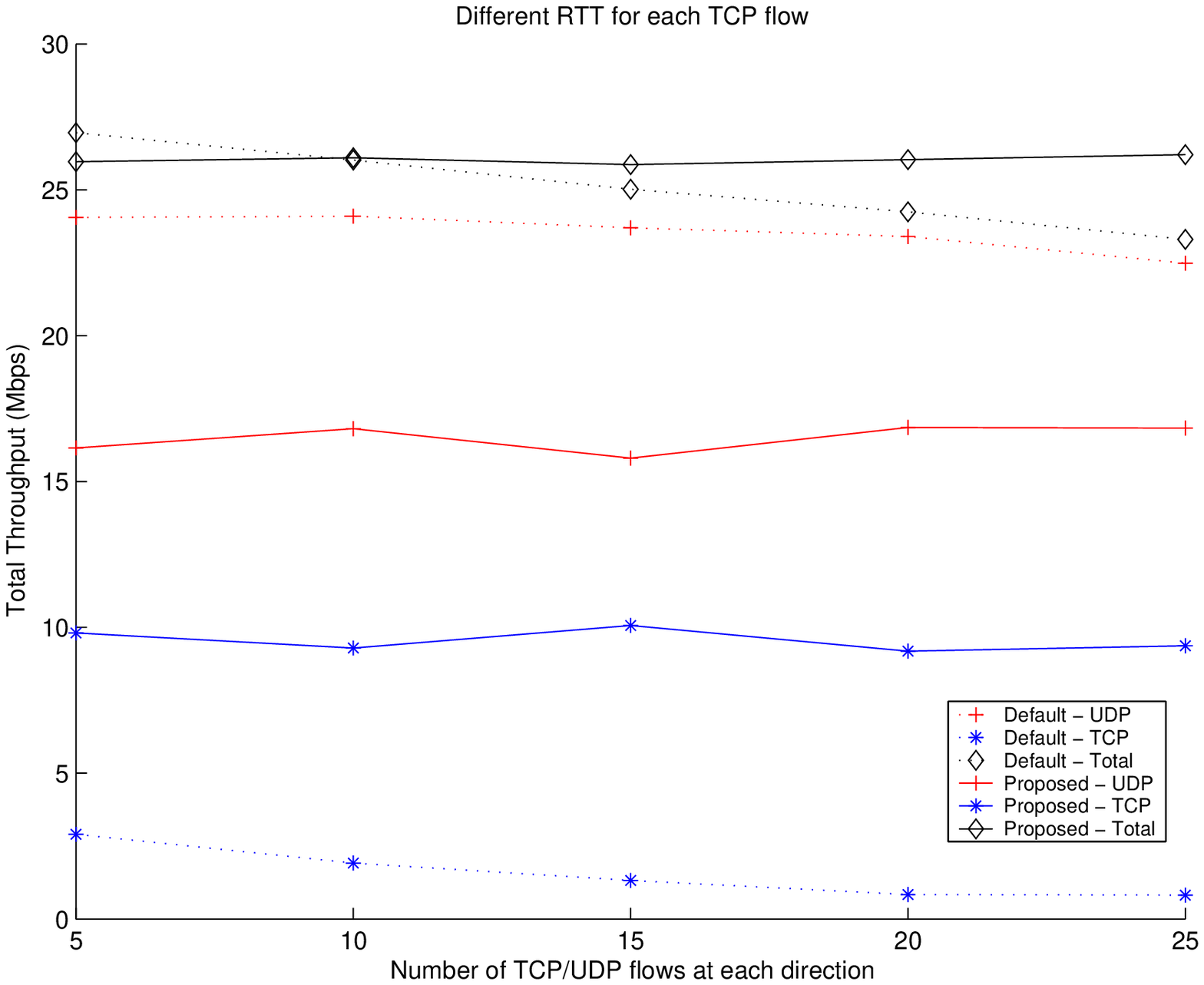}
\caption{The total throughput of TCP and UDP flows as well as the
total system throughput (experiment 2).} \label{fig:udp_sc1_fig6}
\end{figure}

\begin{figure}[t]
\centering \includegraphics[width = 1.0\linewidth]{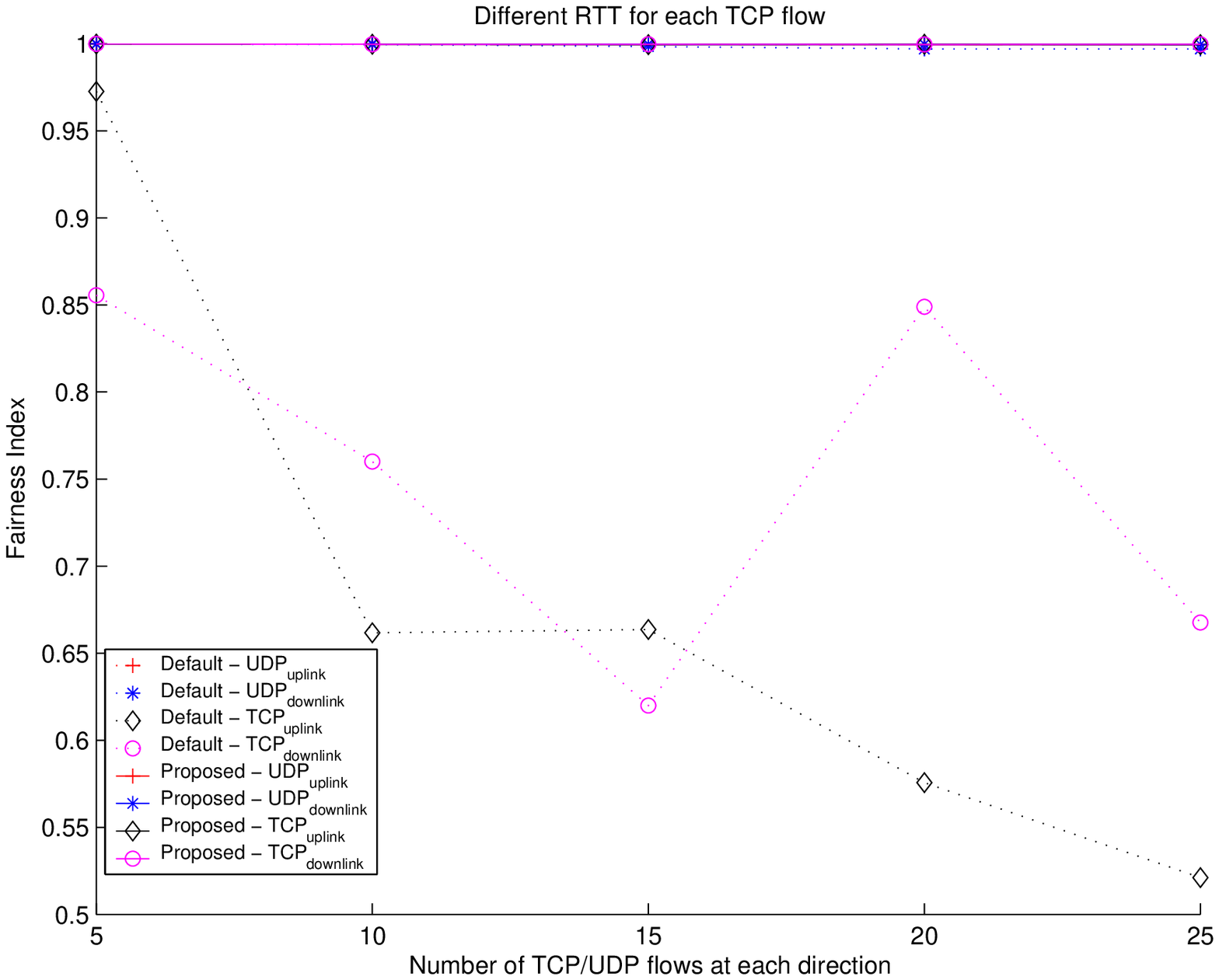}
\caption{Fairness index of individual TCP or UDP flows in the same
direction (experiment 2).} \label{fig:udp_sc1_fig2}
\end{figure}

\paragraph{Experiment 2}
We have repeated the simulation set of experiment 1 when the wired
link delay is varied for TCP flows. The wired link delay of the
first TCP connection is set to 24 ms and each newly generated TCP
connection is assigned 4 ms larger wired link delay than the
previous one. Therefore, the second TCP connection has 28 ms wired
link delay, the third one has 32 ms wired link delay and so on.
This holds for both uplink and downlink connections. UDP wired
link delay is constant for each connection.
Fig.~\ref{fig:udp_sc1_fig4} shows the average throughput of each
TCP and UDP flow in each direction. Fig.~\ref{fig:udp_sc1_fig6}
shows the total throughput of TCP and UDP flows as well as the
total system throughput. Fig.~\ref{fig:udp_sc1_fig2} shows the
performance in terms of fairness between individual TCP or UDP
flows in the same direction. As the results show, the performance
of the proposed algorithm in terms of fair resource allocation is
independent of the duration of the wired link delay. High channel
utilization and perfect fairness is maintained. On the other hand,
as the comparison of Fig.~\ref{fig:udp_sc1_fig1} and
Fig.~\ref{fig:udp_sc1_fig2} imply, the performance of default EDCA
depends on the duration of the wired link delay. In the case of
varying wired link delays, the unfairness between individual TCP
flows both in the downlink and uplink is even worse.

\begin{figure}[t]
\centering \includegraphics[width = 1.0\linewidth]{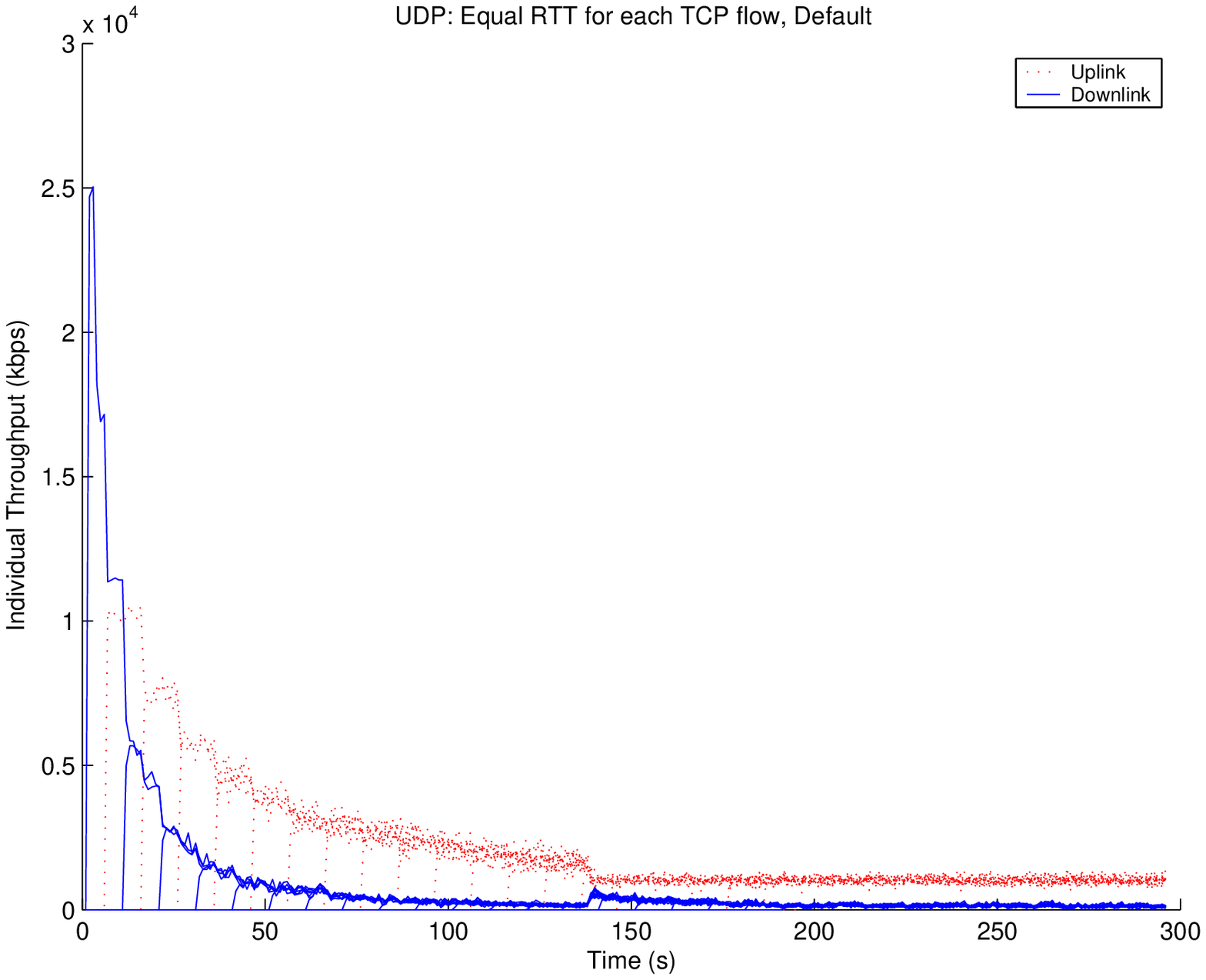}
\caption{The instantaneous UDP throughput of individual uplink and
downlink flows for default EDCA (experiment 3).}
\label{fig:udp_sc2_fig1}
\end{figure}

\begin{figure}[t]
\centering \includegraphics[width = 1.0\linewidth]{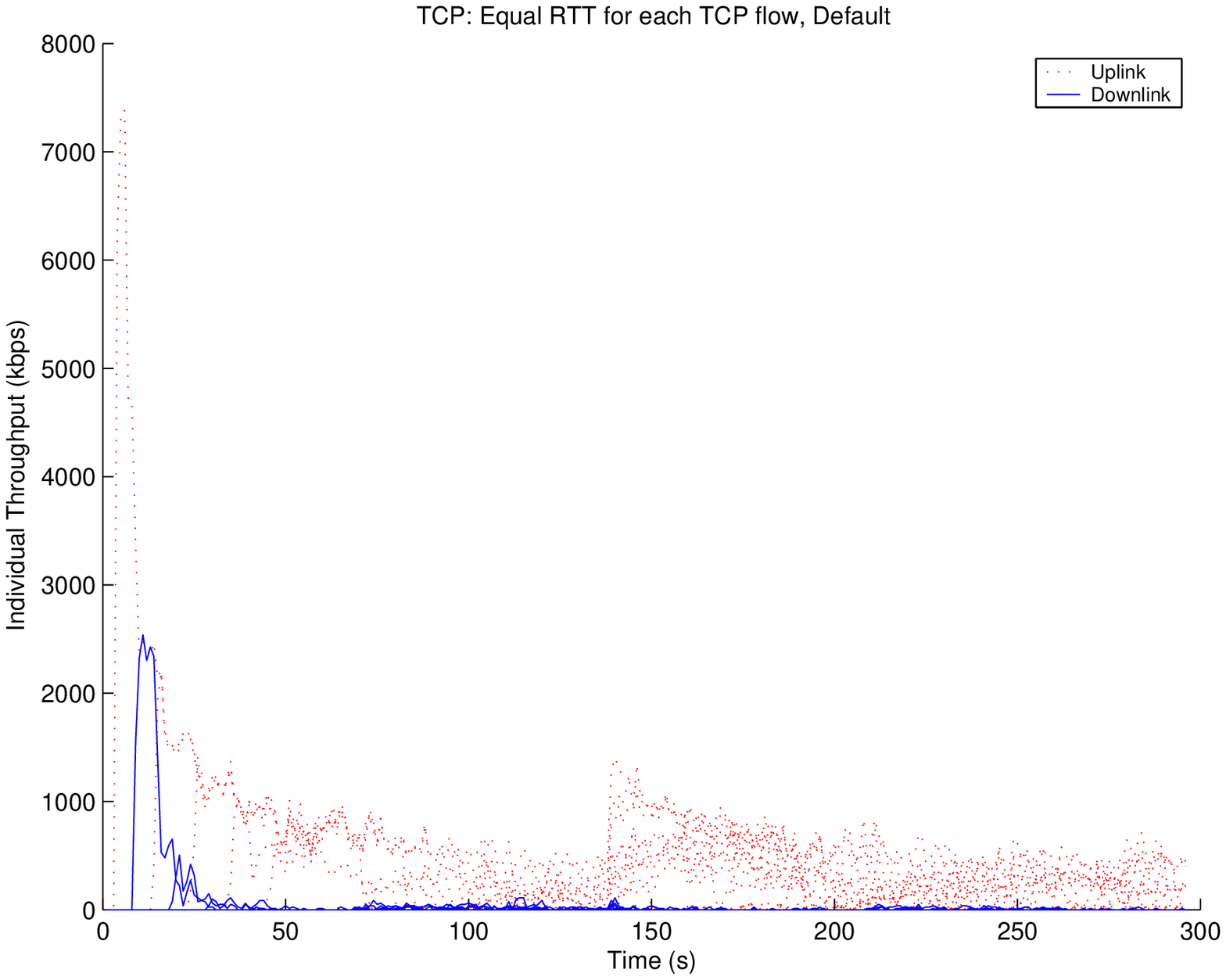}
\caption{The instantaneous TCP throughput of individual uplink and
downlink flows for default EDCA (experiment 3).}
\label{fig:udp_sc2_fig2}
\end{figure}

\begin{figure}[t]
\centering \includegraphics[width = 1.0\linewidth]{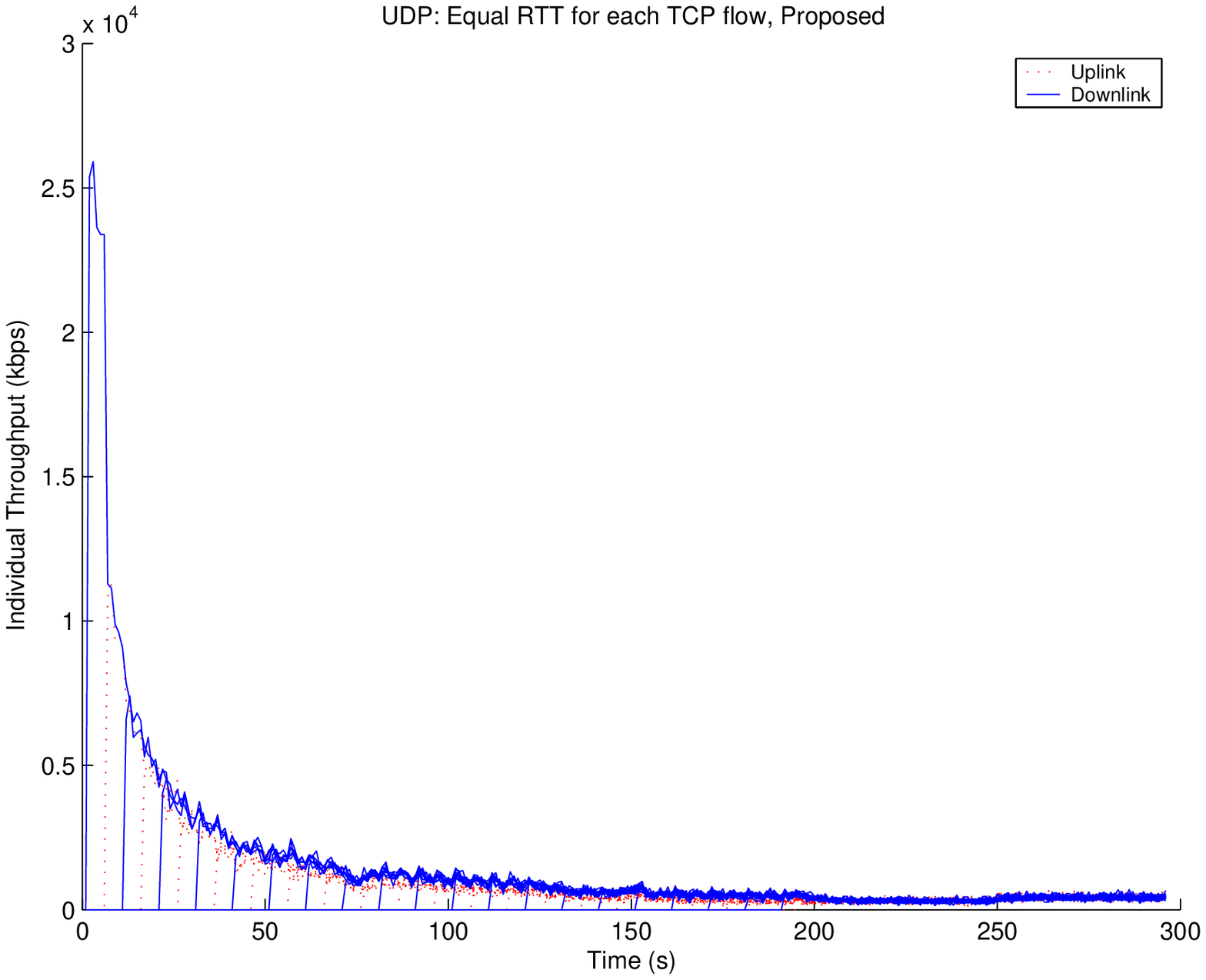}
\caption{The instantaneous UDP throughput of individual uplink and
downlink flows for the proposed algorithm (experiment 3).}
\label{fig:udp_sc2_fig3}
\end{figure}

\begin{figure}[t]
\centering \includegraphics[width = 1.0\linewidth]{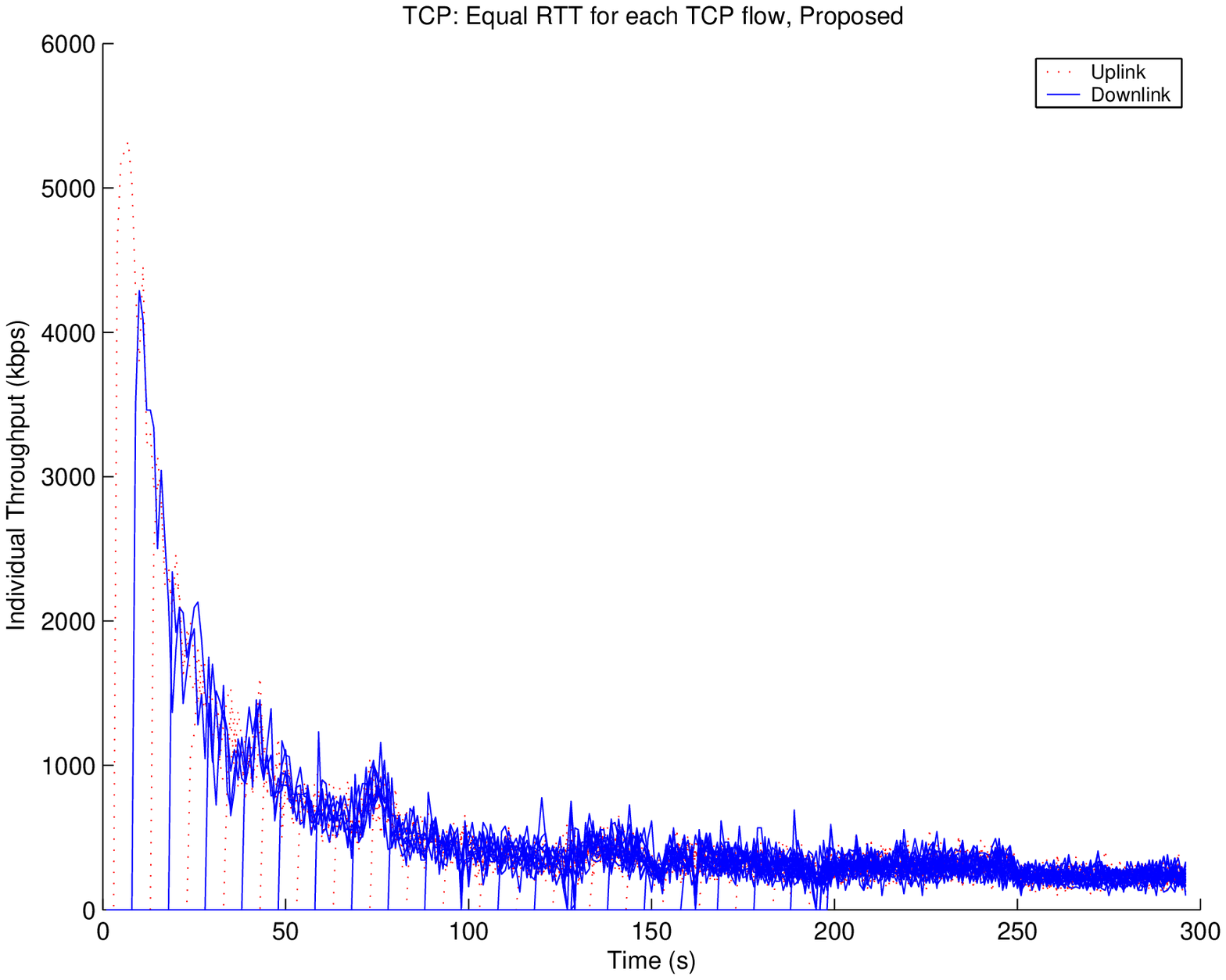}
\caption{The instantaneous TCP throughput of individual uplink and
downlink flows for the proposed algorithm (experiment 3).}
\label{fig:udp_sc2_fig4}
\end{figure}

\paragraph{Experiment 3}
In the third set of experiments, we also generate an equal number
of TCP and UDP flows both in the uplink and downlink. In this
scenario, each uplink or downlink flow starts at different times
and the simulation duration is 300 seconds. The wired link delay
is equal for all flows. The first downlink UDP connection starts
at $t=5$ s. The first uplink UDP connection starts at $t=10$ s.
The first uplink TCP connection starts at $t=7$ s. The first
downlink TCP connection starts at $t=12$ s. Then, a new flow of
the same type arrives every 10 s. No other flow arrives after 200
s. Fig.~\ref{fig:udp_sc2_fig1} and Fig.~\ref{fig:udp_sc2_fig2}
show the instantaneous UDP and TCP throughput of individual uplink
and downlink flows respectively for default EDCA. The unfairness
between uplink and downlink for both UDP and TCP and the
unfairness between individual TCP flows both in the uplink and
downlink are evident. Fig.~\ref{fig:udp_sc2_fig3} and
Fig.~\ref{fig:udp_sc2_fig4} show the instantaneous UDP and TCP
throughput of individual uplink and downlink flows respectively
when the proposed algorithm is enabled. As the results imply, the
proposed algorithm adaptively updates EDCA parameters and always
maintains instantaneous $U_{r}$ (as calculated in (\ref{eq:Ur}).

\begin{figure}[t]
\centering \includegraphics[width = 1.0\linewidth]{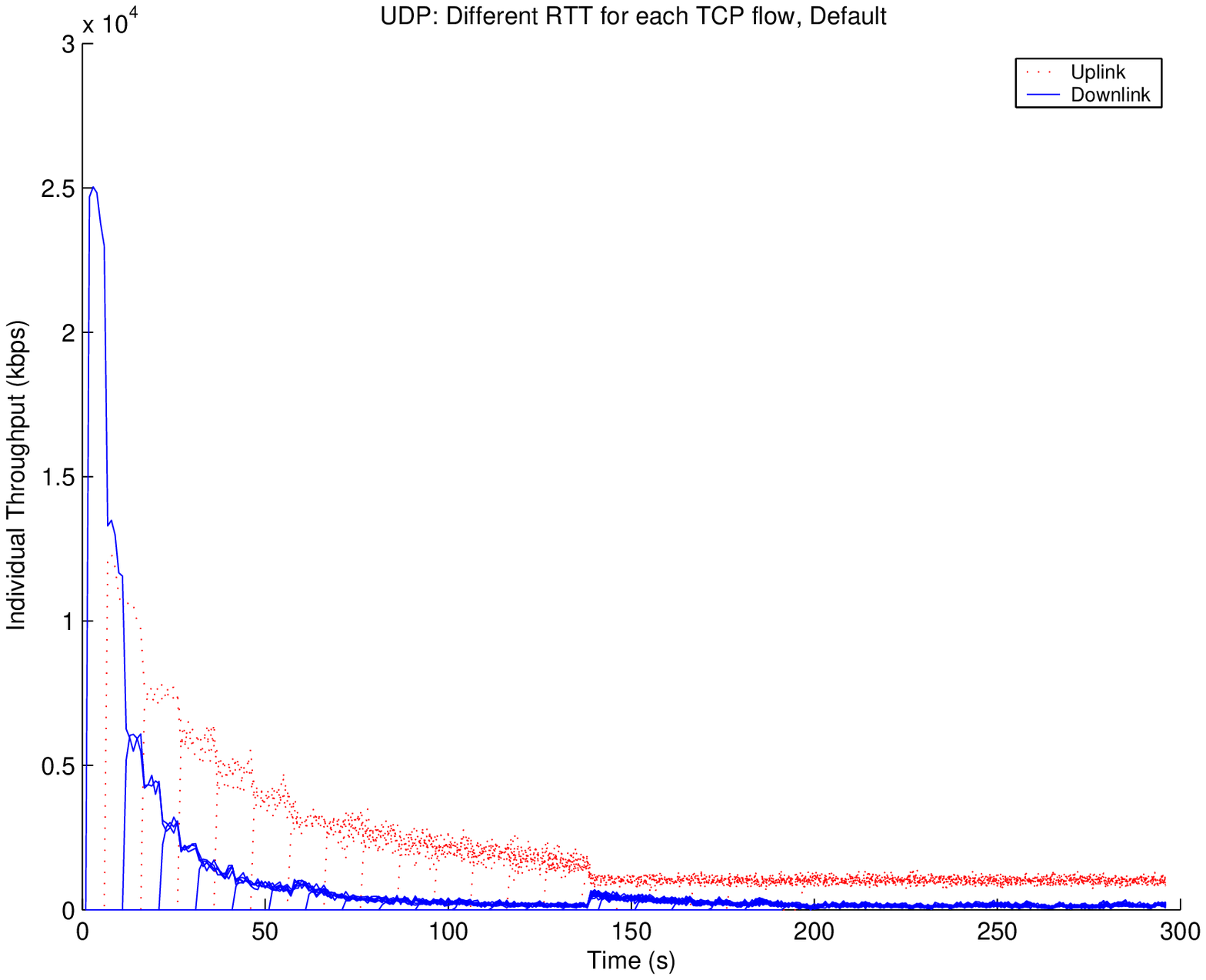}
\caption{The instantaneous UDP throughput of individual uplink and
downlink flows for default EDCA (experiment 4).}
\label{fig:udp_sc2_fig5}
\end{figure}

\begin{figure}[t]
\centering \includegraphics[width = 1.0\linewidth]{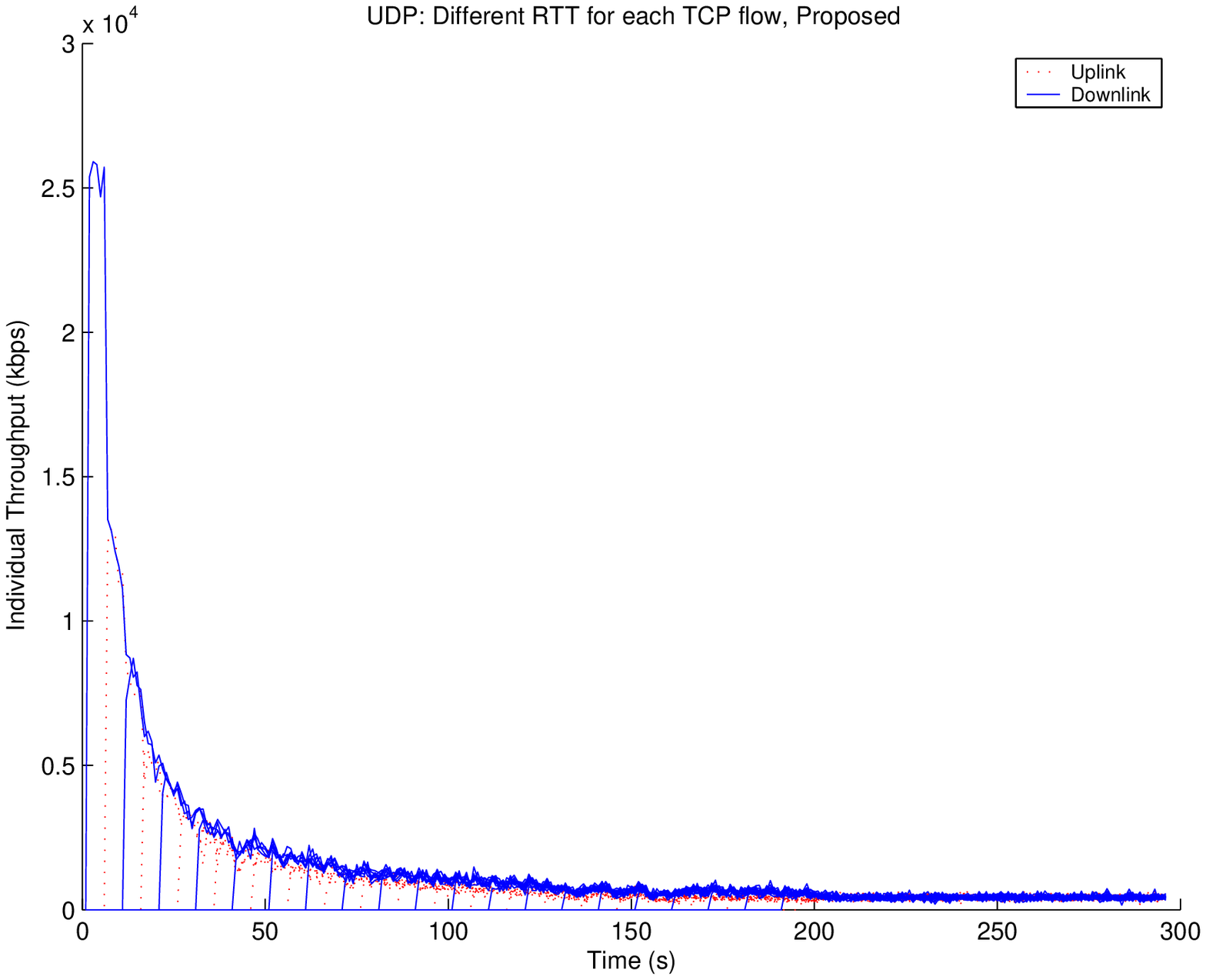}
\caption{The instantaneous TCP throughput of individual uplink and
downlink flows for default EDCA (experiment 4).}
\label{fig:udp_sc2_fig6}
\end{figure}

\begin{figure}[t]
\centering \includegraphics[width = 1.0\linewidth]{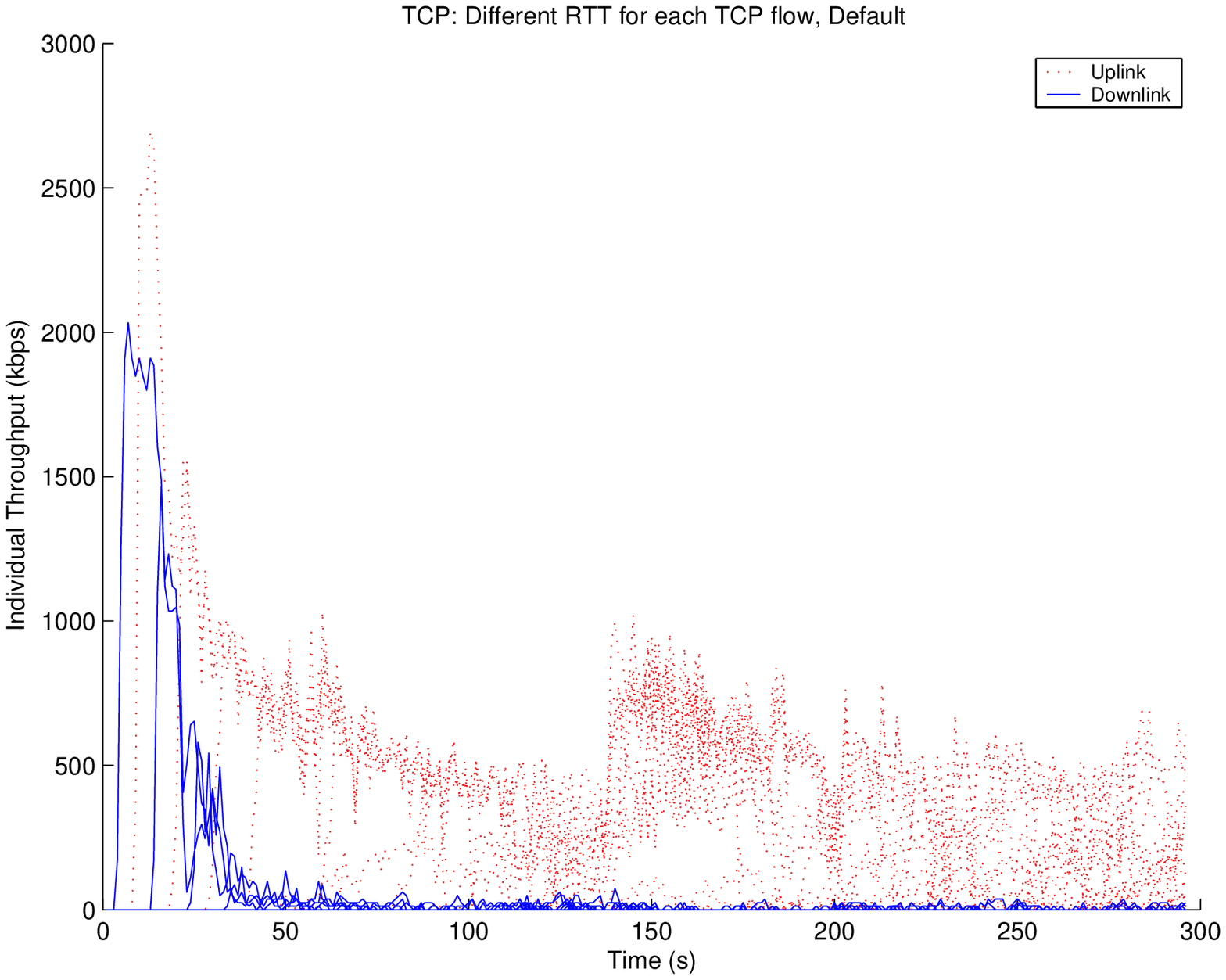}
\caption{The instantaneous UDP throughput of individual uplink and
downlink flows for the proposed algorithm (experiment 4).}
\label{fig:udp_sc2_fig7}
\end{figure}

\begin{figure}[t]
\centering \includegraphics[width = 1.0\linewidth]{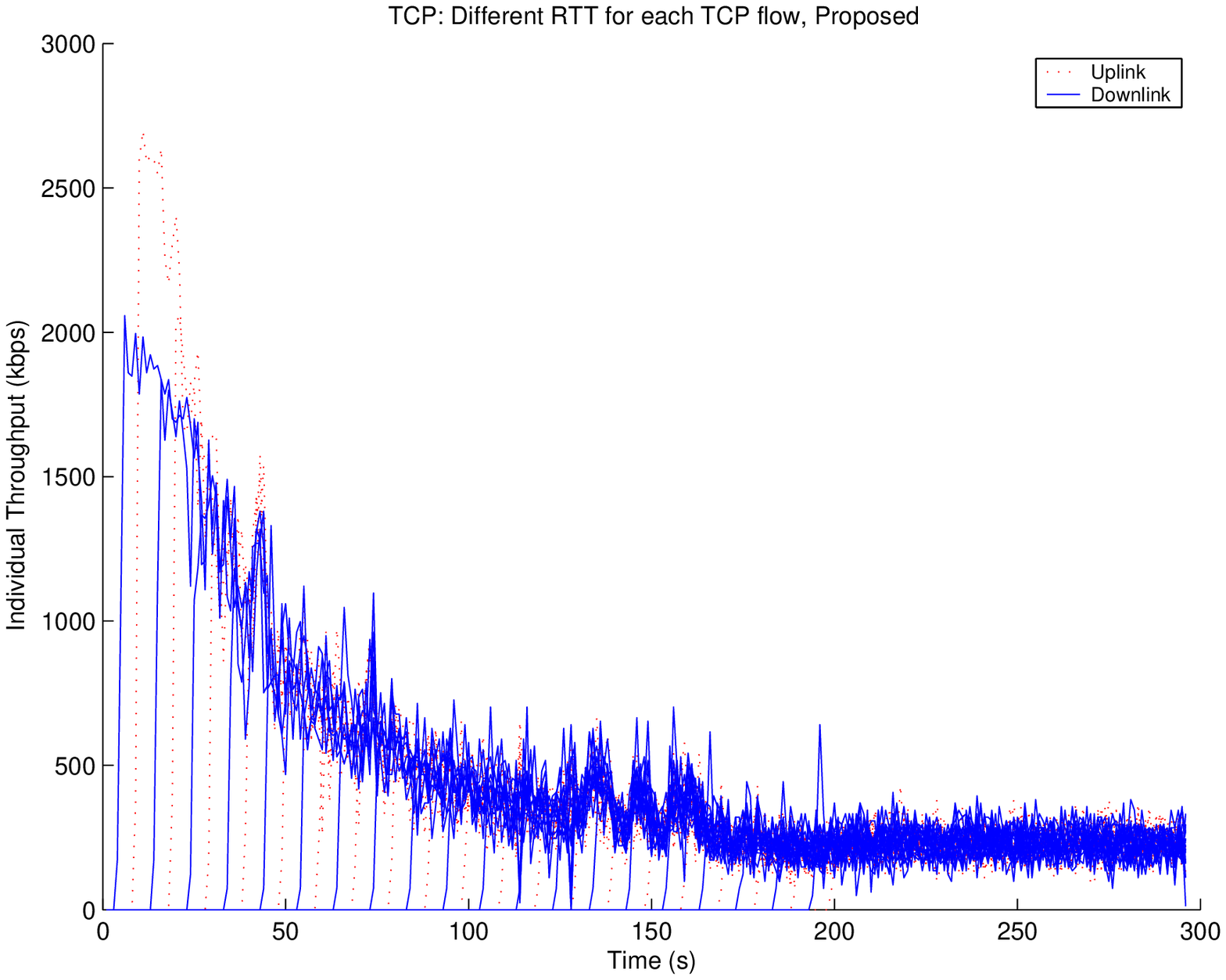}
\caption{The instantaneous TCP throughput of individual uplink and
downlink flows for the proposed algorithm (experiment 4).}
\label{fig:udp_sc2_fig8}
\end{figure}

\paragraph{Experiment 4}
We have repeated the simulation set of experiment 3 when the wired
link delay is varied for TCP flows. We set different wired link
delays using the way as previously stated.
Fig.~\ref{fig:udp_sc2_fig5} and Fig.~\ref{fig:udp_sc2_fig6} show
the instantaneous UDP and TCP throughput of individual uplink and
downlink flows respectively for default EDCA. The unfairness
between individual TCP flows both in the uplink and downlink are
more pronounced when compared with the equal wired link delay
scenario. Fig.~\ref{fig:udp_sc2_fig7} and
Fig.~\ref{fig:udp_sc2_fig8} show the instantaneous UDP and TCP
throughput of individual uplink and downlink flows respectively
for the proposed algorithm. Since the proposed algorithm
adaptively updates EDCA parameters, it maintains fair resource
allocation. The downlink flows does not starve in terms of
throughput.

\paragraph{Experiment 5}

\begin{figure}[t]
\centering \includegraphics[width = 1.0\linewidth]{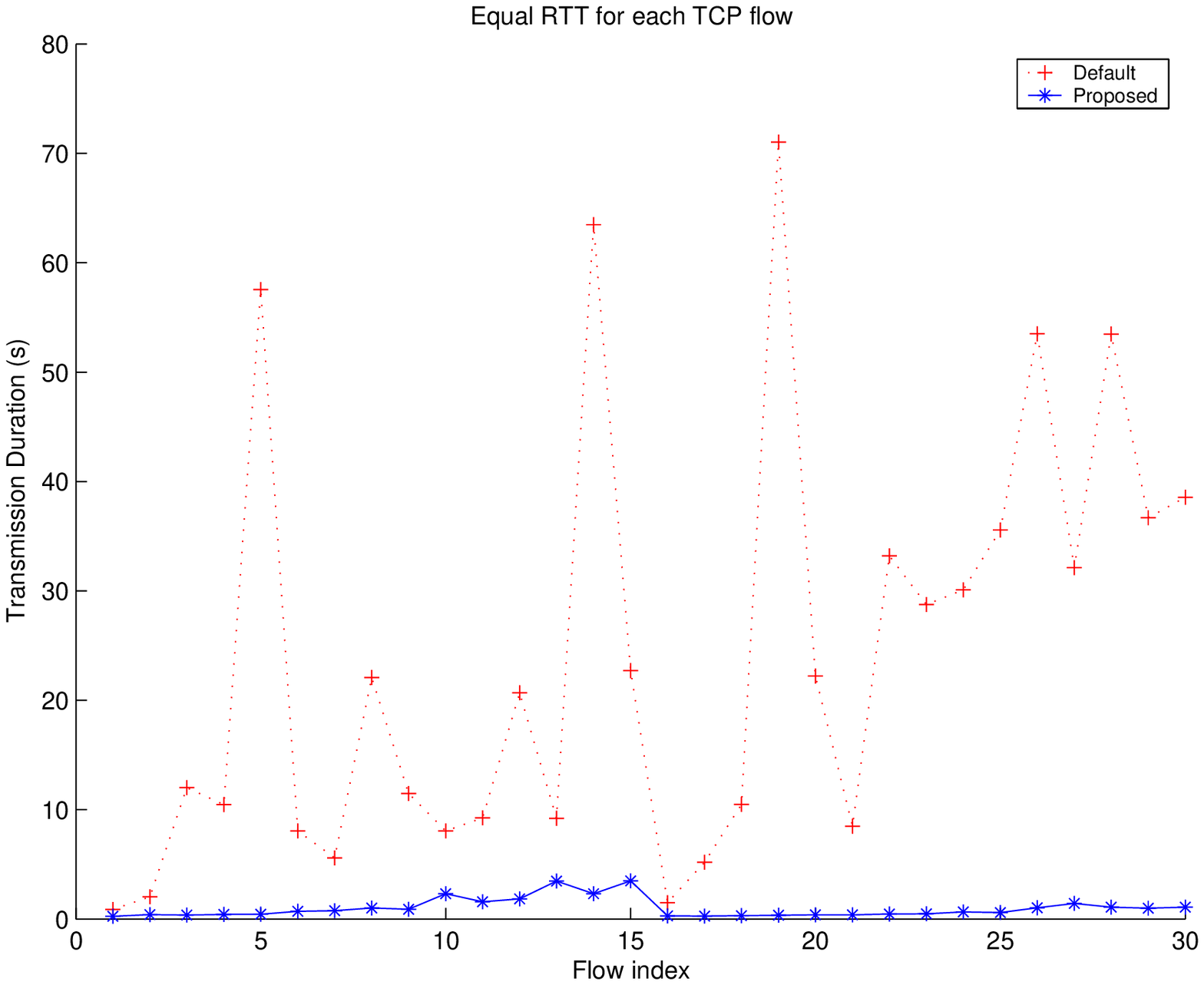}
\caption{The total transmission duration for individual short TCP
flows(experiment 5).} \label{fig:udp_sc3_fig1}
\end{figure}

We have repeated the simulation set of experiment 3 when half of
the TCP flows model short flows. The flow generation times follow
the rules of experiment 3. The simulation duration is 450 s.  No
other flow arrives after 300 s. The short and long TCP flows are
alternatively initiated both in the downlink and uplink. The short
TCP flows consist of 31 packets and leave the system after all the
data is transferred. Fig.~\ref{fig:udp_sc3_fig1} shows the total
transmission duration for individual short TCP flows for the
proposed algorithm and the default EDCA. Note that flow indices
from 1 to 15 represent uplink TCP flows while flow indices from 16
to 30 represent downlink TCP flows. The file transfers with short
durations can be completed in a considerably shorter time when the
proposed algorithm is used. At high load, short flows experience
significantly long delays and connection timeouts when default
constant EDCA parameter selection is used.

\paragraph{Experiment 6}

\begin{figure}[t]
\centering \includegraphics[width = 1.0\linewidth]{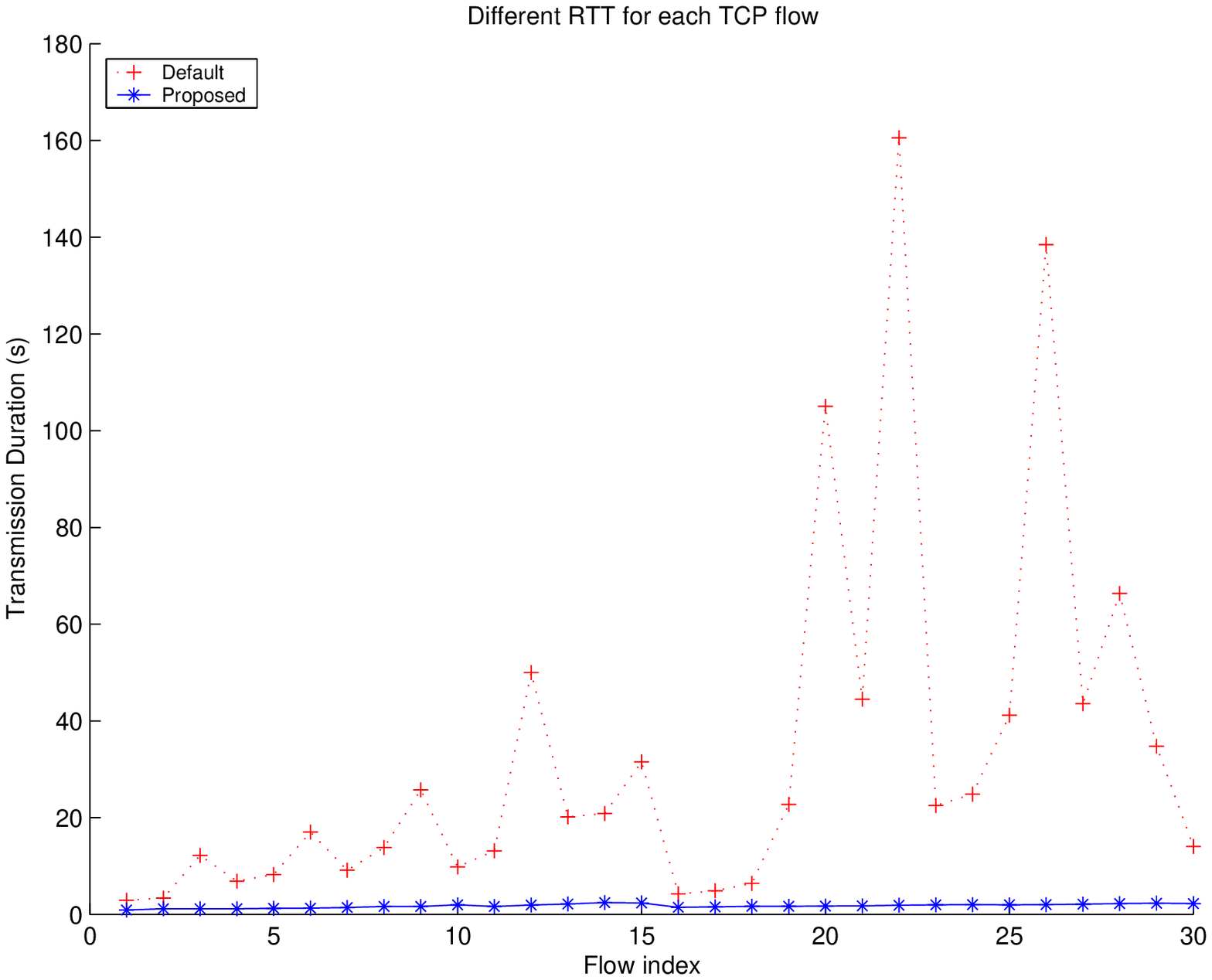}
\caption{The total transmission duration for individual short TCP
flows (experiment 6).} \label{fig:udp_sc3_fig6}
\end{figure}

We have repeated the simulation set of experiment 5 when the wired
link delay is varied for TCP flows. We set different wired link
delays using the way as previously stated.
Fig.~\ref{fig:udp_sc3_fig6} shows the total transmission duration
for individual short TCP flows for the proposed algorithm and the
default EDCA. The comparison of Fig.~\ref{fig:udp_sc3_fig1} and
Fig.~\ref{fig:udp_sc3_fig6} reveals that the proposed algorithm
performance in terms of short TCP flow completion time is
independent of varying wired link delays among the flows.

\begin{figure}[t]
\centering \includegraphics[width = 1.0\linewidth]{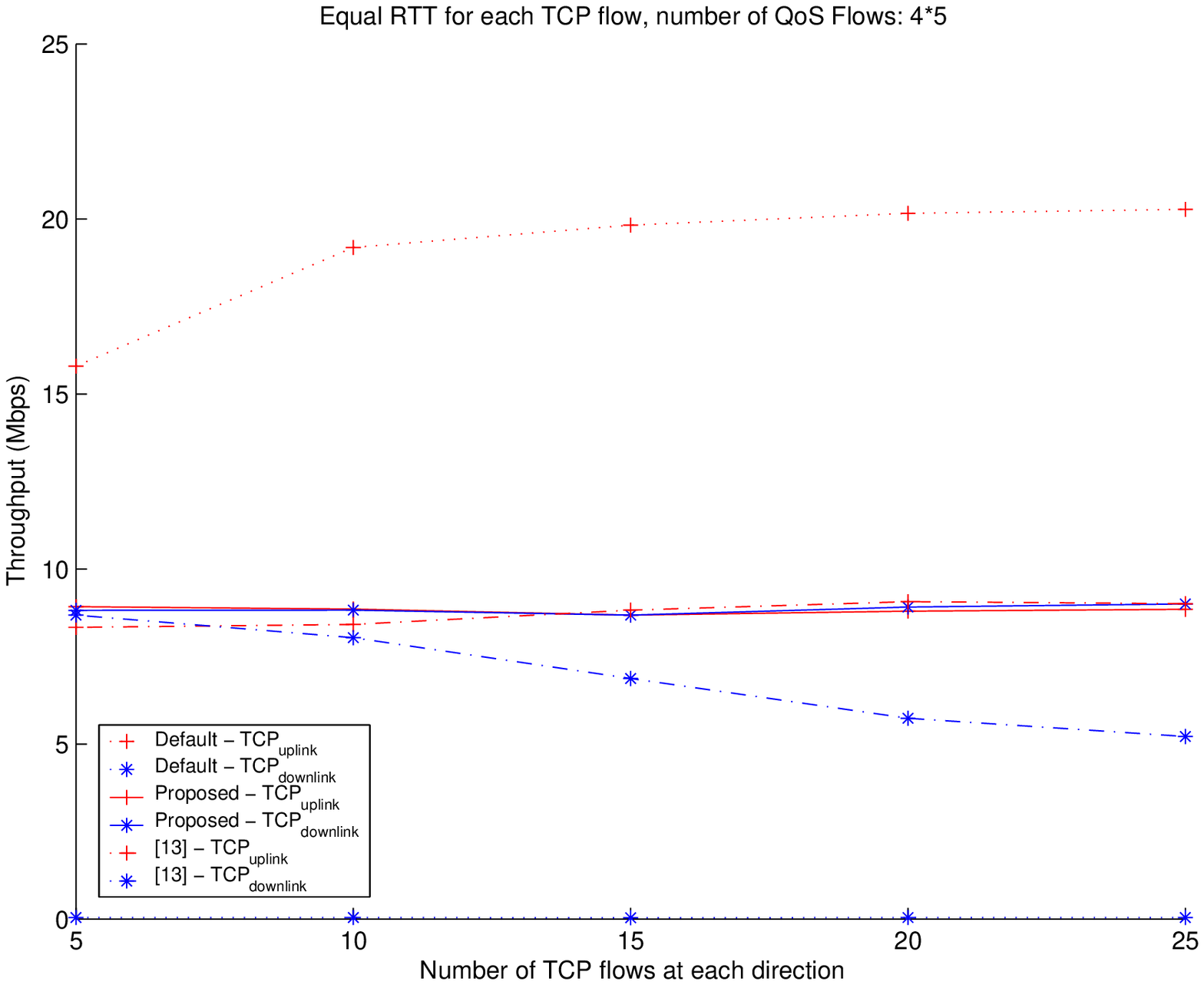}
\caption{The average throughput of uplink and downlink data flows
when there are 5 voice and 5 video flows both in the uplink and
downlink (experiment 7).} \label{fig:qos_sc1_fig4}
\end{figure}

\begin{figure}[t]
\centering \includegraphics[width = 1.0\linewidth]{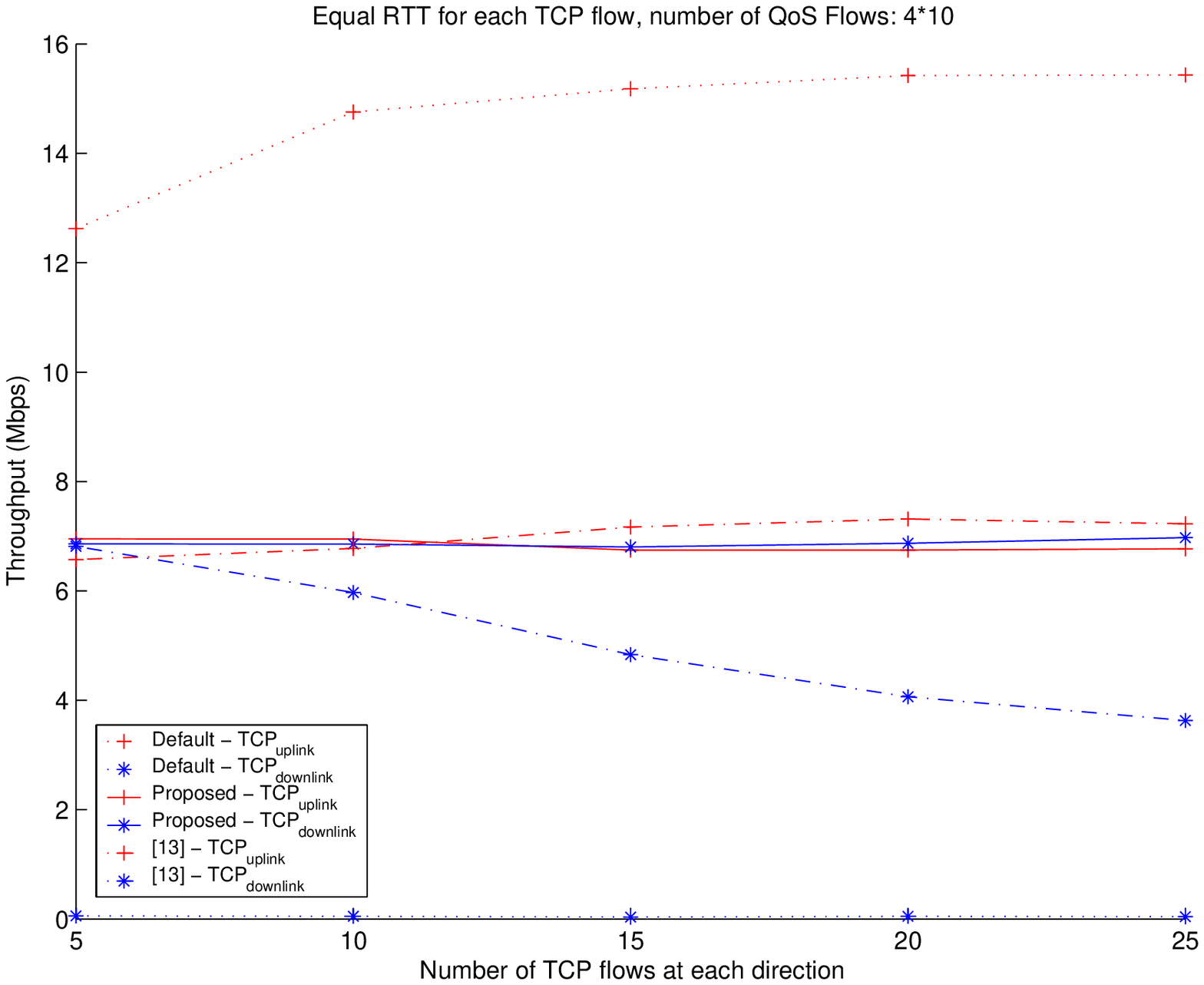}
\caption{The average throughput of uplink and downlink data flows
when there are 10 voice and 10 video flows both in the uplink and
downlink (experiment 7).} \label{fig:qos_sc1_fig5}
\end{figure}

\begin{figure}[t]
\centering \includegraphics[width = 1.0\linewidth]{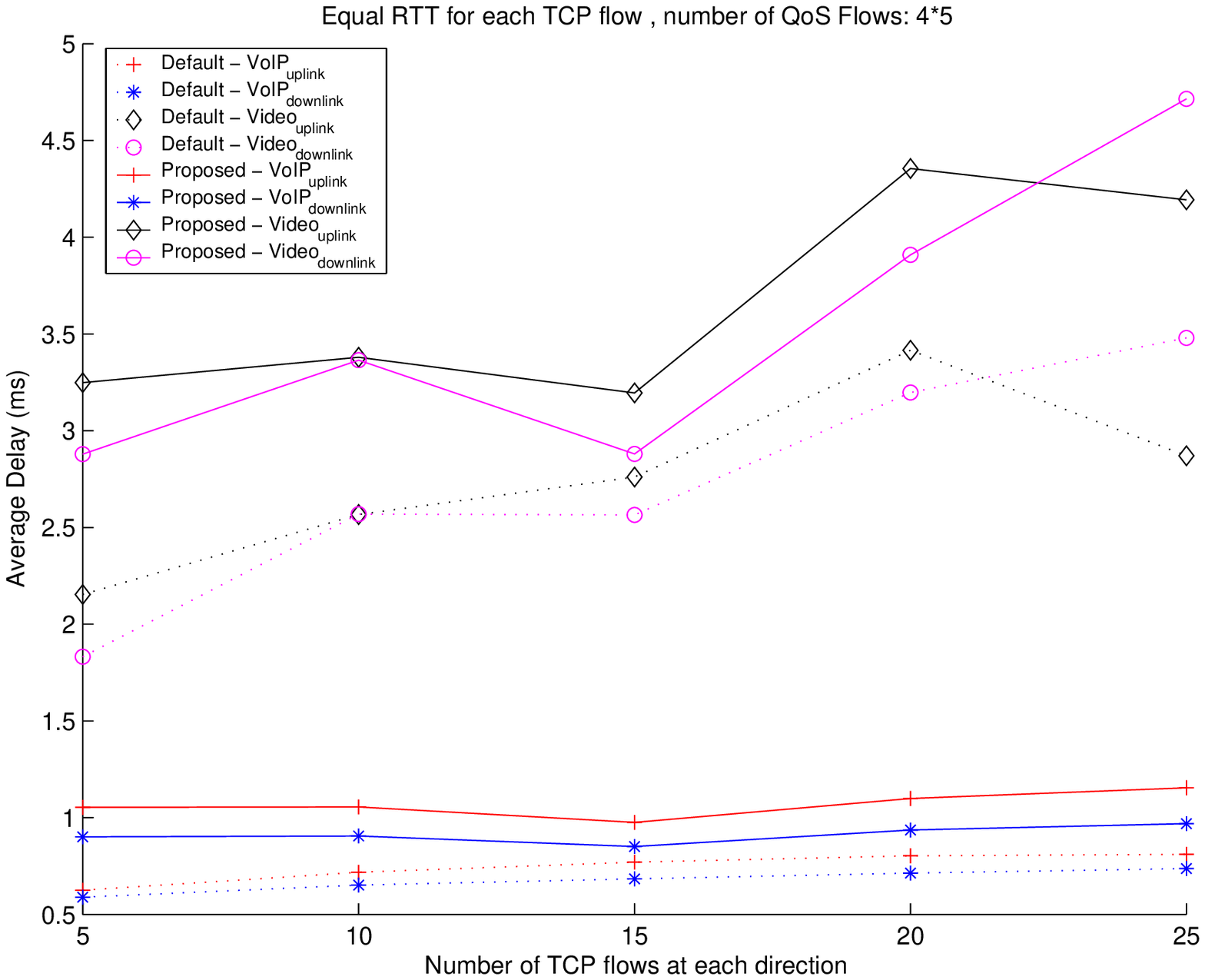}
\caption{The average delay of each QoS flow in each direction when
there are a total of 20 flows with QoS requirements (experiment
7).} \label{fig:qos_sc1_fig7}
\end{figure}

\begin{figure}[t]
\centering \includegraphics[width = 1.0\linewidth]{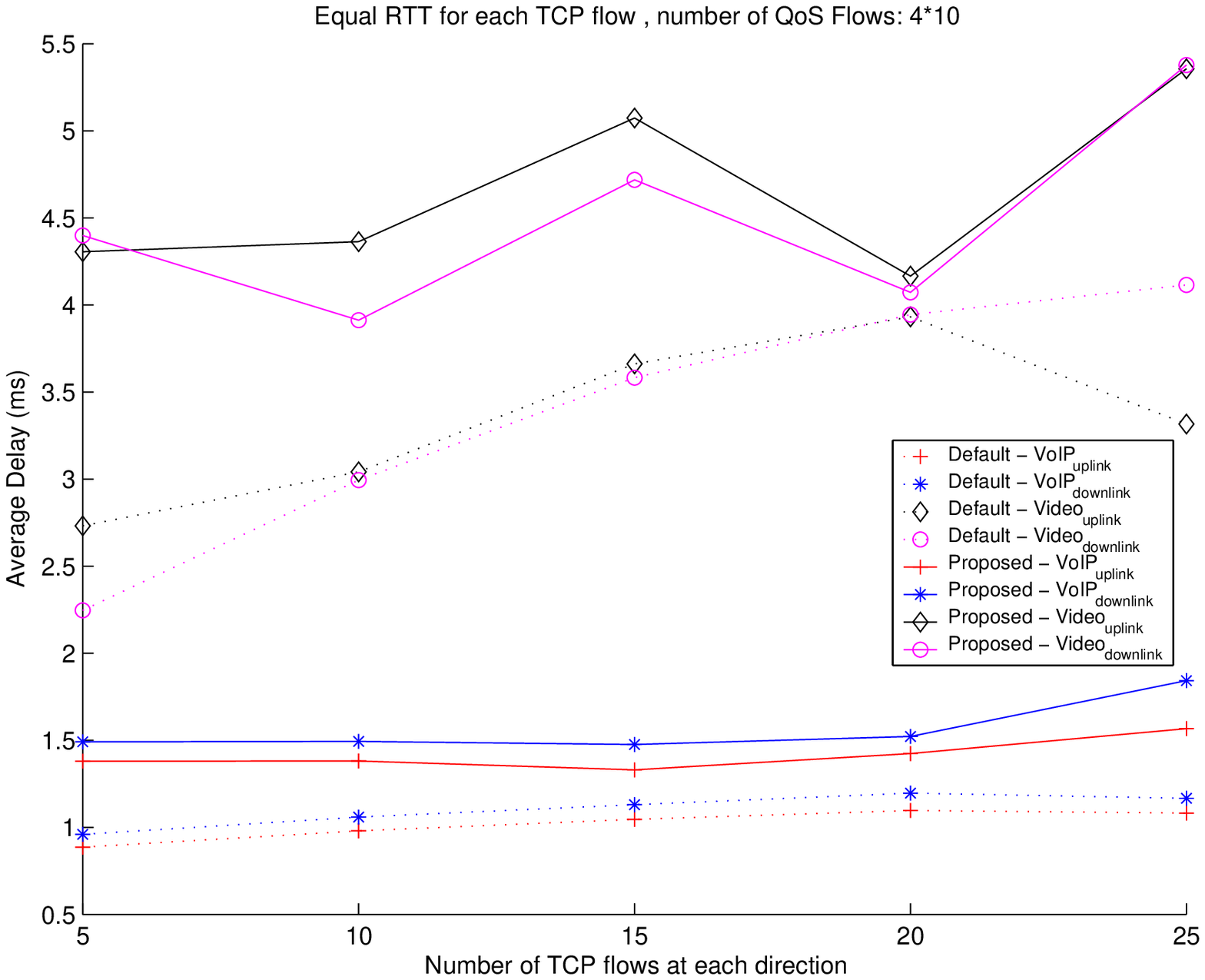}
\caption{The average delay of each QoS flow in each direction when
there are a total of 40 flows with QoS requirements (experiment
7).} \label{fig:qos_sc1_fig8}
\end{figure}

\paragraph{Experiment 7}
In another set of experiments, we consider three types of traffic
sources; audio, video, and data. The audio traffic model
implements a Voice-over-IP (VoIP) application as a Constant Bit
Rate (CBR) traffic profile at 24 kbps. The constant audio packet
size is 60 bytes. Although not presented here, similar results and
discussion hold when the silence suppression scheme is used and
the audio traffic exhibits on-off traffic characteristics. For the
video source models, we have used traces of real H.263 video
streams \cite{Seeling04}. The mean and maximum video payload size
is 2419 bytes and 3112 bytes respectively. The mean video data
rate is 255 kbps. The audio flows are mapped to an AC with
$CW_{min}=7$ and $CW_{max}=15$. The video flows use an AC with
$CW_{min}=15$ and $CW_{max}=31$. For both ACs, AIFSN values are
set to 2 and TXOP limits are 0. Fig.~\ref{fig:qos_sc1_fig4} shows
the average throughput of uplink and downlink data flows when
there are 5 voice and 5 video flows both in the uplink and
downlink (a total of 20 flows with QoS requirements). Similarly,
Fig.~\ref{fig:qos_sc1_fig5} shows the average throughput of uplink
and downlink data flows when there are 10 voice and 10 video flows
both in the uplink and downlink. We also compare the results with
the proposed algorithm of \cite{Leith05}. As the results of
default EDCA and \cite{Leith05} imply, sticking with constant EDCA
parameters for any number of flows does not result in fair access
no matter which EDCA parameter setting is used. On the other hand,
the proposed adaptive algorithm effectively manages fair resource
allocation for any number of stations. Note that we have not
included the average throughput of the flows with QoS requirements
in Fig.~\ref{fig:qos_sc1_fig4} and Fig.~\ref{fig:qos_sc1_fig5},
since all audio and video flows get necessary bandwidth to serve
offered load with zero packet loss rate.
Fig.~\ref{fig:qos_sc1_fig7} compares the average delay of each QoS
flow in each direction for default EDCA and the proposed algorithm
when there are a total of 20 flows with QoS requirements.
Similarly, Fig.~\ref{fig:qos_sc1_fig8} compares the average delay
of each QoS flow in each direction for default EDCA and the
proposed algorithm when there are a total of 40 flows with QoS
requirements. As the results show, the QoS flows experience
slightly larger delays when the proposed algorithm is used (due to
smaller CW and larger TXOP assignment for data flows). On the
other hand, the delay increase is well within the limits of QoS
requirements. Moreover, fair resource allocation for data flows is
provided.

\section{Conclusions}

We have proposed a model-assisted measurement-based dynamic EDCA
parameter adaptation algorithm that achieves a predetermined
utilization ratio between uplink and downlink flows of the same AC
while keeping the prioritization among ACs. The key contribution
is that depending on simple network measures, the proposed
algorithm dynamically adapts the EDCA parameters calculated via a
proposed analytical model. Another key insight is that the
proposed algorithm differentiates the way of adaptation between
UDP and TCP flows regarding their characteristics.
%Up to the
%authors' knowledge, none of the work in the literature on IEEE
%802.11 MAC upload/download fairness considered the asymmetry in
%the forward and backward link packet rate during the slow-start
%phase of the TCP connections.

The proposed algorithm is fully compliant with the 802.11e
standard. We propose AP to use any CW value, not necessarily
exponents of 2. Our observation is that the 802.11e standard does
not restrict the CW settings of the ACs at the AP to be the powers
of 2, while the CW setting of the ACs at the STA should be powers
of 2 due to the definition of specific fields in the beacon
packet. Our approach provides the AP the freedom of satisfying any
required utilization ratio through fine tuning on CW settings.

Via simulations, it is shown that fair resource allocation between
uplink and downlink flows of an AC can be maintained in a
wide-range of scenarios when the proposed model-assisted
measurement-based dynamic EDCA parameter adaptation algorithm is
used. The performance of the proposed algorithm in terms of fair
resource allocation is shown to be independent of the duration of
the round trip time of a connection. Short flows experience
significantly low delays and no connection timeouts. Therefore, we
conclude that the proposed method also provides short-term
fairness. The QoS requirements of existing audio and video flows
in the 802.11e WLAN are maintained. Our results also show that
sticking with constant EDCA parameters at any scenario does not
result in fair access no matter which EDCA parameter setting is
used.

%GATHER{C:/INANCINAN/bibliography/standards.bib}
%GATHER{C:/INANCINAN/bibliography/HCCA.bib}
%GATHER{C:/INANCINAN/bibliography/simulations.bib}
%GATHER{C:/INANCINAN/bibliography/channel.bib}
%GATHER{C:/INANCINAN/bibliography/books.bib}
%GATHER{C:/INANCINAN/bibliography/EDCAanalysis.bib}
%GATHER{C:/INANCINAN/bibliography/mypapers.bib}
%GATHER{C:/INANCINAN/bibliography/fairness.bib}
%GATHER{C:/INANCINAN/bibliography/myreports.bib}
%GATHER{EDCAfairness.bbl}

\bibliographystyle{IEEEtran}
\bibliography{IEEEabrv,C:/INANCINAN/bibliography/standards,C:/INANCINAN/bibliography/HCCA,C:/INANCINAN/bibliography/simulations,C:/INANCINAN/bibliography/channel,C:/INANCINAN/bibliography/books,C:/INANCINAN/bibliography/EDCAanalysis,C:/INANCINAN/bibliography/mypapers,C:/INANCINAN/bibliography/fairness,C:/INANCINAN/bibliography/myreports}

\clearpage

\clearpage

\clearpage

\clearpage

\clearpage

\clearpage

\clearpage

\clearpage

\clearpage

\clearpage

\clearpage

\clearpage

\clearpage

\clearpage

\clearpage

\clearpage

\clearpage

\clearpage

\clearpage

\clearpage

\clearpage

\clearpage

\clearpage

\clearpage

\clearpage

\end{document}